\documentclass{article}
\usepackage{epsfig}
\usepackage{amssymb}
\usepackage{amsbsy}
\usepackage{mathabx}
\usepackage[colorlinks=true, pdfstartview=FitV, linkcolor=blue, citecolor=blue, urlcolor=blue]{hyperref}
\usepackage{verbatim}
\def\wt{\widetilde}
\def\wc{\widecheck}
\def\wh{\widehat}
\def\cs{\widetilde}
\def\ds{\displaystyle}

\def\res{\mathop{\mathrm{res}}\limits_}

\makeatletter
\@addtoreset{equation}{section}
\makeatother
\def\tr{\mathrm {Tr}}
\def\&{&{\hskip -20pt}}
\def\m{\mathop}
\def \p{\mathbf p}
\def \s{\mathbf s}

\def\1{{\bf 1}}
\def\un{\underline}
\def \pa{\partial}
\def\C{{\mathbb C}}
\def\R{{\mathbb R}}
\def\N{{\mathbb N}}
\def\Z{{\mathbb Z}}
\def\ds{\displaystyle}
\newtheorem{theorem}{Theorem}[section]
\newtheorem{coroll}{Corollary}[section]
\newtheorem{lemma}{Lemma}[section]
\newtheorem{remark}{Remark}[section]
\newtheorem{proposition}{Proposition}[section] 
\newtheorem{definition}{Definition}[section]
\def\ri{\right}				\def\le{\left}
\def\1{{\bf 1}}
\def\Amat{{\widehat{\mathbb A}}}
\def\br{\begin{remark}}		\def\er{\end{remark}}
\def\bt{\begin{theorem}}		\def\et{\end{theorem}}
\def\bc{\begin{coroll}}		\def\ec{\end{coroll}}
\def\bl{\begin{lemma}}		\def\el{\end{lemma}}
\def\bd{\begin{definition}}		\def\ed{\end{definition}}
\def\bp{\begin{proposition}}	\def\ep{\end{proposition}}
\def\be{\begin{equation}}		\def\ee{\end{equation}}
\def\bea{\begin{eqnarray}}	\def\eea{\end{eqnarray}}
\def \pa{\partial}
\def\Int{\int\!\!\!\int_\varkappa}
\def\A{\mathop{\mathbf a}}

\def\L{\mathcal L}
\def\Wp{{\ds \mathop {\Psi}_{\infty}}}
\def\Ws{{\ds \mathop{\Phi}_{\infty}}}
\def\unWs{{\ds \mathop{\un\Phi}_{\infty}}}
\def\a{{\alpha}}
\def\b{{\beta}}
\date{}
\begin{document}
\fontfamily{cmss}
\fontsize{10pt}{15pt}
\selectfont



\begin{titlepage}
\begin{flushright}
CRM-3205 (2005)\\
\end{flushright}
\vspace{0.2cm}
\begin{center}
\begin{Large}\fontfamily{cmss}
\fontsize{17pt}{27pt}
\selectfont
\textbf{ Biorthogonal polynomials for  2-matrix models with semiclassical potentials}
\end{Large}\\
\bigskip
\begin{large} {M.
Bertola}$^{\dagger\ddagger}$\footnote{Work supported in part by the Natural
    Sciences and Engineering Research Council of Canada (NSERC),
    Grant. No. 261229-03 and by the Fonds FCAR du
    Qu\'ebec No. 88353.}\footnote{bertola@crm.umontreal.ca}
\end{large}
\\
\bigskip
\begin{small}
 $^{\ddagger}$ {\em Department of Mathematics and
Statistics, Concordia University\\ 7141 Sherbrooke W., Montr\'eal, Qu\'ebec,
Canada H4B 1R6} 
\\
\smallskip
$^{\dagger}$ {\em Centre de recherches math\'ematiques,
Universit\'e de Montr\'eal\\ C.~P.~6128 succ. centre ville, Montr\'eal,
Qu\'ebec, Canada H3C 3J7
}\\
\end{small}
\end{center}
\bigskip
\newpage
\begin{center}{\bf Abstract}\\
\end{center}
We consider the biorthogonal polynomials associated to the two--matrix
model where the eigenvalue distribution has  potentials $V_1,V_2$ with arbitrary rational
derivative and whose supports are  constrained on an arbitrary union of intervals
(hard-edges). We show that these polynomials satisfy certain
recurrence relations with a number of  terms $d_i$  depending on the number
of hard-edges and on the degree of the rational functions $V_i'$.
Using these relations we derive Christoffel--Darboux identities
satisfied by the biorthogonal polynomials: this enables us to give explicit formul\ae\ for the differential equation satisfied by $d_i+1$ consecutive polynomials, 
We also define certain integral transforms of the polynomials and use
them to formulate a Riemann--Hilbert problem for $(d_i+1) \times (d_i+1)$ matrices constructed out of the polynomials and these transforms.
Moreover we prove that the Christoffel--Darboux pairing can be interpreted as a
pairing between two dual Riemann--Hilbert problems. 
\medskip
\begin{small}
\end{small}
\bigskip
\bigskip
\bigskip
\bigskip
\end{titlepage}

\tableofcontents

\section{Introduction and setting}
In this paper we consider the biorthogonal polynomials associated to the two--matrix model.
The model is defined by a measure on the space of pairs of Hermitean matrices $M_1,M_2$ of the form  
\be
{\rm d} \mu (M_1,M_2):=  {\rm d} M_1{\rm d}M_2 {\rm e}^{ -\tr(V_1(M_1)-V_2(M_1) + M_1M_2) }\ .
\ee
Using Itzykson--Zuber/Harish-Chandra's formula  the model can be reduced to the study of
biorthogonal polynomials \cite{eynardmehta} (BOPs for short), namely two sequences of polynomials $\{\pi_n(x)\},\{\sigma_n(y)\}$
\bea
\int_\R\int_\R {\rm d}x {\rm d}y {\rm e}^{-V_1(x)-V_2(y)+xy}\pi_n(x) \sigma_m(y) = \delta_{nm}\ .
\eea
For the model to have a probabilistic interpretation, the potentials should be real and satisfy certain growth conditions to ensure the convergence of the integrals.
In order to introduce the setting of this paper we consider the following situation (which is strictly included in the more general setting to be expounded later)
\begin{enumerate}
\item There is a finite collection of disjoint intervals $I = \bigcup I_j\subseteq \R_x$ and $J = \bigcup_j J_j\subseteq \R_y$ ($\R_x$ denotes the real axis of the $x$-variable), in the complement of which the potentials are $+\infty$: in other words the matrices $M_1,M_2$ have spectrum confined to these multi-intervals, so that the associated BOPs  satisfy
\be
 \int_I\int_J{\rm d}x {\rm d}y {\rm e}^{-V_1(x)-V_2(y)+xy}
 \pi_n(x) \sigma_m(y) = \delta_{nm}
\ee
\item The two potentials $V_1(x)$ and $V_2(y)$  are the restriction to
  $I,J$ (respectively) of  real-analytic functions with rational
  derivative (with poles symmetrically placed off the real axis, or on
  the complement of the intervals on the real axis)  together with the necessary growth condition if the intervals are unbounded.
\end{enumerate}

This situation has been addressed in \cite{jat} within the general context of bilinear moment functionals. Indeed it is convenient to recast the orthogonality condition in a more abstract setting where one considers a {\bf bimoment functional} $\L:\C[x]\otimes \C[y]\to \C$ defined by 
\be
\L(x^i | y^j) := \int_I\int_J {\rm d}x {\rm d}y \, x^i y^j {\rm e}^{-V_1(x)-V_2(y)+xy} =\mu_{ij}\ .\label{therealthing}
\ee
and then extended by linearity to arbitrary polynomials.
The biorthogonality condition then  reads
\be
\L(\pi_n| \sigma_m) = \delta_{nm}\ .
\ee
The properties of the potentials $V_1,V_2$ and the supports of
integration can be dealt with on the same footing by purely algebraic
methods: to this end one introduces four polynomials $A_i, B_i$,
$i=1,2$ according to the strategy outlined hereafter. 
Let $(x_j,m_j)$ be the location of the poles of $V_1'(x)$ with their
order (we include all of the poles, in this case also the complex
conjugates, which clearly come in with the same multiplicities) and let
$a_j$ be the endpoints of $I$. We define then $A_1,B_1$ (and similar
expressions for $A_2,B_2$) as follows 
\be
B_1(x) = \prod (x-x_j)^{m_j} \prod (x-a_j)\ ,\qquad
A_1 := V_1' B_1 - B_1'\ ,
\ee
so that now $V_i' = \frac {A_i + B_i'}{B_i}$. 
It is a straightforward exercise to verify (using integration by
parts) that the bimoment functional satisfies the following
distributional  identities for arbitrary $p(x)\in \C[x], s(y)\in \C[y]$ 
\bea
&& \L\Big(-B_1(x)p'(x)+ A_1(x) p(x)\Big|s(y)\Big) =
\L\Big(B_1(x)p(x)\Big|ys(y)\Big) \label{semi1}\ ,
\\
&&\L\Big(p(x)\Big|-B_2(y) s'(y) +A_2(y) s(y)\Big) =
\L\Big(xp(x)\Big|B_2(y)s(y)\Big)\label{semi2}\ .
\eea
Abstracting formul\ae\ (\ref{semi1},\ref{semi2}) from the specific
context, we will say that a bimoment functional $\L$ is {\bf
  semiclassical} if it satisfies those same relations
(\ref{semi1},\ref{semi2}) for some given (and fixed) polynomials
$A_i,B_i$. The name comes from a similar usage in the context of
ordinary orthogonal polynomials \cite{marcellan}.
 
Such functionals have been studied in \cite{jat}, where it was shown that 
\bp
For given $A_i,B_i$, $i=1,2$, a semiclassical moment functional $\L$
is the linear combination of $s_1s_2$ independent functionals
$\L_{\nu,\mu},\ \mu=1\dots,s_1,\ \nu=1,\dots,s_2$, where $s_i =
\max(\deg A_i,\deg B_i+1)$ 
\ep
More importantly (at least in the case $\deg A_i\geq \deg B_i+1$) all
of these moment functionals $\L_{\mu,\nu}$ can be given an integral
representation completely analog to (\ref{therealthing}),  but without
any restriction on the  reality of the potentials or of the contours of
integration: this is the setting of the present paper.  
%
\subsection{Connection to other orthogonal polynomials}
The algebraic properties  of semiclassical bilinear moment functionals apply to a slightly different class of orthogonal polynomials. Let us consider in fact orthogonal polynomials in the complex plane with respect to a measure of the form 
\be
{\rm d}\mu (z,\overline z) := {\rm e }^{-|z|^2 +2 \Re V(z)}{\rm d}^2 z 
\ee
where $V(z)$ is a holomorphic function such that $V'(z)$ is rational. The convergence of the measure mandates that the residues of $V'(z){\rm d}z$ must have real part greater than $-\frac 1 2 $ and that the behavior at $\infty$ of $V$ cannot exceed the second power (and also a certain open  condition on the coefficient of this quadratic term  which we do not specify here). 
Orthogonal polynomials are defined as a holomorphic basis of $L^2(\C,{\rm d}\mu)$. It is amusing to note that the moment functional 
\be
\L(z^j | \overline z^k) := \iint_{\C} z^{i}\overline z^k{\rm d}\mu(z ,\overline z) = :\mu_{j\overline k }
\ee
is a semiclassical moment functional (using Stokes' thm. {\em in vece} of integration by parts) with just some (obvious) reality constraint on the bimoments. 
Therefore all the algebraic manipulations that rely on the
semiclassicity alone carry out verbatim to this case and confirm
certain manipulations used in \cite{razvan}. With very minor and trivial modifications,  Section \ref{semibi}
almost entirely generalizes (in particular Thm. \ref{thmrecmul}).   
Significant differences (sufficient to require a different analysis to appear elsewhere) arise in the construction of the fundamental systems and the Riemann--Hilbert problem.

\subsection{Connection to $2$-Toda equations}
The framework of this paper is connected to  the general theory of $2$-Toda equations \cite{UT,AvM, AvM2}. This
is the  theory of a pair of (semi)-infinite matrices $P,Q$ (in our
notation) where $Q$ is lower-Hessenberg and $P$ is
upper--Hessenberg\footnote{We say that a matrix is lower Hessenberg its
  $(i,i+1+k)$ entries vanish ($\forall k=1,2,\dots$) and also  all
  $(i,i+1)$-entries are nonzero. A matrix is upper Hessenberg if its
  transposed is lower-Hessenberg.} which  evolve under a bi-infinite set
of commuting flows $\{t_j,\wt t_j\}_{j\in \N}$  
\bea
&& \pa_{t_J}Q = -\frac 1 J[Q,(Q^J)_{-0}]\ ,\qquad \pa_{\wt t_J} Q = -\frac 1 J [Q,(P^J)_{-0}]\\
&&\pa_{t_J}P =-\frac 1 J  [P,(Q^J)_{+0}]\ ,\qquad \pa_{\wt t_J} P = -\frac 1 J [P,(P^J)_{+0}]
\eea
where the subscript $_{\pm 0}$ denotes the upper/lower triangular part plus half of the diagonal (we are assuming the normalization such that the upper triangular part of $Q$ coincides with the transposed of the lower-triangular part of $P$).

Let now $Q,P$ be semi-infinite matrices. 
We can  use $Q, P$ to denote the matrices expressing the multiplicative recurrence relations of a sequence of polynomials,
\be
x\pi_n = \sum_{j=0}^{n+1} Q_{nj} \pi_j\ ,\qquad 
y\sigma_n = \sum_{j=0}^{n+1} P_{nj} \sigma_n\ ,
\ee
where the polynomials are recursively {\em defined} by this relation.
 Using the generalization of  Favard's theorem proved in our
 \cite{jat} we prove  the existence of (unique) a bimoment functional $\L:\C[x]\otimes \C[y]\to \C$ such that 
 \be
 \L(\pi_n| \sigma_m) = \delta_{nm}\ .
 \ee 
It then follows easily that the $2$-Toda  flows are {\bf linearized} by this moment  map, in the sense that the solutions $Q({\bf t},\wt {\bf t}), P({\bf t},\wt {\bf t})$ are simply the multiplication matrices for the biorthogonal polynomials of the moment functional
\be
\L_{{\bf t},\wt {\bf t}} (\bullet | \bullet ) := \L\le({\rm e}^{-\sum \frac {t_J}J x^J} \bullet \big| {\rm e}^{ - \sum \frac {\wt t_J}{J}y^J} \bullet \ri)\ .
\ee
The moment functionals of semiclassical type (eqs. \ref{semi1},
\ref{semi2}) that we are going to analyze form a particular class of
    {\bf reductions} of the above-mentioned 2-Toda hierarchy. The
    simplest situation is the one of bimoment semiclassical
    functionals with polynomial potentials as the ones considered in
    \cite{BEH, BEH2, BEH3}, where the matrices $P,Q$ are also {\em finite
      band}. Moreover the solutions which arise in the context of
    semiclassical bilinear functionals also satisfy the (compatible)
    constraint of the {\bf string equation} 
\be
[P,Q]=\hbar \1
\ee
(the constant $\hbar$ can be disposed of by a rescaling).
The parameters of the (finite--dimensional) reduction are the
coefficients of the potentials: for more general semiclassical moment
functionals as the ones considered in this paper, the parameters
involve not only the coefficients of the partial fraction expansions
of the (derivatives of the) potentials, but also the position of the
poles and the position of the end-points of the supports of the
measure (the hard--edge endpoints).

The paper is organized as follows
\begin{enumerate}
\item In Section \ref{semibi} we derive the recurrence relation satisfied by
the biorthogonal polynomials of a semiclassical moment
functional. There are two types of recurrence relations: one which
involves the multiplication by the spectral parameter (and plays the
r\^ole of the more standard three--term recurrence relation for
orthogonal polynomials) and one which involves a differential operator
acting on the polynomials.
\item In Section \ref{bilconco} we recall some possibly not well
  known facts about a certain class of linear homogeneous ODEs. These
  equations are next in simplicity to the class of constant
  coefficients ODEs, inasmuch as the coefficients are allowed to be
  linear functions of the independent variable. When considering the
  formal adjoint equation then the classical bilinear
  concomitant provides a nondegenerate pairing between the solution
  spaces of the pair of mutually adjoint ODEs. In this case we give an
  interpretation of it in terms of an {\bf intersection pairing}
  between certain contours used in the  representation of  the
  solutions as contour-integrals. This part of the paper is
  logically quite independent on the rest but it is nevertheless necessary
  in order to understand certain constructs of  the following section.
\item In Section \ref{auxwave} we define the auxiliary wave vectors
  for our functionals, using a certain multiple integral transform
  which relies upon the form of the bilinear concomitant associated to
  our semiclassical moment functional (extending some of the  results of \cite{BHI}). These expression will prove
  crucial in the formulation of a first order ODE of rank $d_i + 1 = 1+ 
  \deg(A_i)$ satisfied by the biorthogonal polynomials.
  We also derive the analog of the Christoffel--Darboux identities
  satisfied by standard orthogonal polynomials to our case of
  biorthogonal polynomials: similar expression were extensively used in
  \cite{BEH2,BHI} for the case where the potentials $V_i$ are
  polynomials (which is a subcase  strictly included in our present
  setting) and in absence of hard-edge endpoints. The novel feature is
  that these new identities involve not only the biorthogonal
  polynomials of the moment functional $\L$ itself, but also those of
  the {\bf associated} bilinear semiclassical moment functionals 
  \be
   \wc \L:= \L(B_1 \bullet | \bullet)  \ ;\qquad \wh \L:= \L(\bullet |
  B_2\bullet)\ . 
  \ee 
  This feature appears prominently in the {\em perfect duality} of the
  Riemann--Hilbert problems  appearing in the next section.
\item In Section \ref{dualRH} we define a pair\footnote{In fact there
  are two such pairs, the other being obtained by interchanging the
  r\^oles of $x,y$, $B_1,B_2$ etc.} of piecewise--analytic
  matrices constructed out of the entries of the wave-vectors and
  their auxiliary wave-vectors. They satisfy certain jump conditions
  on contours in the complex plane and some asymptotic behavior at
  the zeroes of $B_1$. Moreover they satisfy rational first order ODEs with poles
  at the zeroes of $B_1$. The Christoffel--Darboux identity, when
  written as a bilinear expression for these matrices becomes a
  {\em perfect pairing} (Thm. \ref{perfect}) in the sense that
  establishes a nondegenerate constant (in $x$) duality-pairing
  between the two solution spaces. This pairing  should be thought
  of as the ``dressed'' form of the bilinear concomitant pairing
  introduced in Sect. \ref{bilconco}. Similar Riemann--Hilbert
  problems have appeared elsewhere in the literature,
  e.g. \cite{KM,Kap,BHI, BEH}.
\end{enumerate}
In order to facilitate the navigation through the paper all proofs of
 more technical nature are
collected in the appendix and only those that may help the understanding
are left in the main body of the paper. 

{\bf \large Acknowledgments.} The author would like to thank R. Teodorescu for discussion during the summer 2005 conference on Random Matrices at CRM, John Harnad for daily stimulating interaction and one of the referees for pointing out some mistakes in an early version.

\section{Semiclassical bilinear moment  functionals of type $BB$}
\label{semibi}
We consider an arbitrary bilinear semiclassical moment functional (as
defined in the introduction)  \cite{jat}, i.e. satisfying
(\ref{semi1}, \ref{semi2}). 
Let $q_i=\deg (B_i)$ and $d_i=\deg (A_i)$:  we assume
that $d_i\geq q_i+1$ ("type BB" in the terminology of \cite{jat}). 
We also make the assumption that the two pairs of polynomials
$A_i,B_i$ are {\bf reduced} in the sense that the only common zeroes
of $A_i$ and $B_i$ ($i=1,2$) are amongst the  simple zeroes of $B_i$.
 Any moment functional coming from a
representation like the one in the introduction (\ref{therealthing})
has this property of reducedness. 
In \cite{jat} the case of non-reduced moment functional is also
considered, and it corresponds to functionals which may be expressed
as delta functions (or derivatives thereof): we refer ibidem for
details. 

It is known \cite{jat} that any such reduced  moment functional can be expressed in integral form 
\bea
&&\mu_{ij}:= \L(x^i|y^j) =\sum_{\mu=1}^{d_1} \sum_{\nu=1}^{d_2}
\varkappa_{\mu,\nu} \L_{\mu,\nu} (x^i | y^j) \label{moments}\\ 
&&\L_{\mu\nu} (x^i| y^j) = \int_{\Gamma_{x,\mu}} \int
_{\Gamma_{y,\nu}} {\rm e}^{-V_1(x)-V_2(y)+xy}{\rm d}x {\rm d}y\\ 
&& V_i'(y) = \frac {A_i + B_i'}{B_i}\label{2-3}\\
&& \L= \Int x^iy^j{\rm e}^{-V_1(x)-V_2(y)+xy}{\rm d}x {\rm d}y\\
\eea
The two sets of  contours of integration $\Gamma_{x,\mu}$ and
$\Gamma_{y,\nu}$ are defined in the $x$ and $y$ complex planes
respectively and in completely parallel fashion: we will
define them in  Section \ref{defcont}. We have also introduced the short-hand  notation 
\be
\Int := \sum_{\mu=1}^{d_1} \sum_{\nu=1}^{d_2} \varkappa_{\mu,\nu}\int_{\Gamma_{x,\mu}} \int _{\Gamma_{y,\nu}}
\ee
Note that the case of hard-edges is included: the hard-edges are the zeroes of $B_i$ 
that cancel with the zeroes of the denominator defining $V_i'$ in eq.  (\ref{2-3}).

The constants $\varkappa_{\mu,\nu}\in \C$ are arbitrary (not all
zero). In the paper we will often invoke "genericity" conditions for
the moment functional $\L$: by this we mean that the genericity is in
the choice of the $\varkappa$-constants and not in the choice of
$A_i,B_i$ which we consider as given once and for all. 
All of the genericity conditions that we will use can be translated
into the nonvanishing of certain infinite sequences of minors of the
matrix of bimoments $M = [\mu_{ij}]$: since the moments $\mu_{ij}$ are
linear in $\varkappa$ as per (\ref{moments}), this genericity boils
down to avoiding an at-most-denumerable collection of divisors of
homogeneous polynomials in the $\varkappa$-space. 
\subsection{Biorthogonal polynomials}
Let us consider the biorthogonal polynomials associated to this
bilinear moment functional, namely two sequences of monic polynomials
satisfying the following conditions 
\bea
\{\pi_n(x),\sigma_n(y)\}_{n\in\N}\cr
\pi_n(x) = x^n + \mathcal O(x^{n-1})\cr
\sigma_n(y) = y^n + \mathcal O(y^{n-1})\cr
\L(\pi_n| \sigma_m) = h_n\delta_{nm}\ .
\eea
The existence of these BOPs is guaranteed provided that the principal
minors of the matrix of bimoments do not vanish 
\be
\Delta_n[\mathcal L]:= \det[\mu_{ij} ]_{0\leq i,j,\leq n-1} \neq 0 \qquad \forall n\in \N\ ,
\ee
which also guarantees that $h_n\neq 0\ ,\forall n\in \N$ (\cite{jat}). 
We find it more convenient to deal with the normalized BOPs;
\be
p_n:= \frac {\pi_n}{\sqrt{h_n}}\ ,\qquad s_n:= \frac {\sigma_n}{\sqrt{h_n}}
\ee
We will use the following {\bf quasipolynomials}
\bea
\psi_n:= p_n{\rm e}^{-V_1(x)}\ ,\qquad \phi_n:= s_n {\rm e}^{-V_2(y)}
\eea
and the following semi-infinite vectors ({\bf wave vectors}) 
\bea
\p(x):= [p_0,p_1,\dots,p_n,\dots]^t\ ,\qquad \s(y):= [s_0,s_1,\dots, s_n,\dots]^t\\
\Wp:=\p(x){\rm e}^{-V_1(x)} \ ,\qquad \Ws:=\s(y){\rm e}^{-V_2(y)} 
\eea
It will become necessary to consider the following {\bf associated} semiclassical functionals defined by the relations
\bea
&& \wh \L(p|s):= \L(p|B_2\,s)\ ,\qquad \wc \L(p|s):= \L(B_1\,p|s)\ .
\eea
We leave to the reader the simple check that these are also
semiclassical moment functionals where the potentials are replaced --respectively-- by 
\bea
\wh \L \ \longleftrightarrow \ \le\{\begin{array}{l}
 \wh V_1(x) = V_1(x)\\
 \wh V_2(y) := V_2(y) - \ln B_2(y)
 \end{array} \ri.\\[12pt]
\wc \L \ \longleftrightarrow \ \le\{\begin{array}{l}
\wc V_1(x) := V_1(x) -\ln B_1(x)\\
\wc V_2(y) = V_2(y)
\end{array}
\ri.
\eea
These definitions amount to $\wh A_2 = A_2-B_2',\ \wh B_2 = B_2$ so that $V_2'  = \frac {\wh A_2+ \wh B_2'}{B_2} = \frac {A_2}{B_2}$.

Note, however, that they are defined along the same contours as $\L$ and with the same coefficients $\varkappa$'s.
\subsection{Multiplicative recurrence relations}
We now prove
\bt
\label{thmrecmul} 
The BOPs satisfy the following finite-term recurrence relations:
\bea
&& x\bigg(p_n + \sum_{j=1}^{q_2} \ell_j(n) p_{n-j} \bigg) = \sum_{j=-1}^{d_2} \a_{j}(n) p_{n-j}\label{eq1}\\  
&& y\bigg(s_n + \sum_{j=1}^{q_1} m_j(n) s_{n-j}
\bigg) =  \sum_{j=-1}^{d_1}\b_{j}(n) s_{n-j}\\
&& q_i  = \deg(B_i),\ d_i = \deg(A_i) \nonumber \ ,  
\eea
where $\ell_j(n) = 0$ for $n\leq d_2$ and $m_j(n)=0$ for $n\leq d_1$, under the  genericity assumption
(to be further discussed  in Remark \ref{remark:genericity}) 
\be
\Delta_{n,2}:=
\det\le[
\begin{array}{cc|ccc}
\mu_{10} \ \dots & \mu_{1,q_2-1} & \mu_{00}& \dots & \mu_{0,n-q_2-1} \\
\mu_{20}\ \dots& \mu_{2,q_2-1} & \mu_{10} & \dots & \mu_{1,n-q_2-1} \\
\vdots &&&& \vdots\\
\vdots &&& & \vdots\\
\mu_{n,0}\ \dots &\mu_{n,q_2-1} & \mu_{n-1,0}& \dots & \mu_{n-1,n-q_2-1}
\end{array}
\ri]\neq 0 \  ,\ \ \forall n > q_2\in \N.\label{gen2}
\ee
The coefficients $\a_{-1}(n)$ and $\b_{-1}(n)$ are nonzero for any $n$;
furthermore, under the same genericity assumptions letting $a_i,b_i$ be the leading coefficients of $A_i,B_i$ we have  
\bea
b_2\alpha_{d_2}(n)\sqrt{h_{n-d_2} }= a_2 \ell_{q_2}(n) \sqrt{h_{n-q_2}}
\neq 0\ ,\ \ n\geq d_2\cr
b_1\beta_{d_1}(n)\sqrt{h_{n-d_1} }= a_1 m_{q_1}(n) \sqrt{h_{n-q_1}}
\neq 0\ ,\ \ n\geq d_1
\label{nonzero}
\eea
 \et

{\bf Proof}.
We prove only one relation, the other being proved by interchanging
the r\^oles.\\
The statement  $\alpha_{-1}(n)\neq 0$ follows from the form of the recurrence relation by comparison of the leading coefficients, which gives
\be
\a_{-1}(n) = \sqrt{\frac {h_{n+1}}{h_n}}\neq 0\ .
\ee
The fact that $\ell_j(n)=0$ for $n\leq d_2$ is a choice
of convenience: indeed, since $d_2>q_2$ any $xp_n$ can be written as
a linear combination of the same BOPs of degrees $m=0,\dots, n+1$ for
$n\leq d_2$. \\
Consider $xp_n(x)$: by "integration by parts"  (i.e. using relation
\ref{semi2} from right to left), we immediately conclude that  
\be
xp_n(x) \ \perp\  B_2(y)\,\C\{1,y,\dots, y^{n-d_2-1}\} =: V_n^{(2)}
\ee
Therefore $V_{n-q_2}^{(2)}$ is in the common annihilator of $x p_n(x),\dots, x p_{n-q_2}(x)$.
We now show that it  is generically  possible to fix the coefficients $\ell_j(n)$ of
a linear combination as the left hand side of eq. (\ref{eq1}) such that the result is
perpendicular to {\em any} polynomial $q(y)$ of degree
$\deg(q)<n-d_2$. Let  
\bea
q(y) = B_2(y) a(y) + b(y)
\eea
be the long division of $q$ by $B_2$ with remainder $b$: then
\bea
\L\bigg(xp_n (x)\bigg|  q(y)\bigg ) = \L\bigg(xp_n(x)\bigg|B_2(y) a(y) + b(y)\bigg) = \L\bigg(xp_n(x)\bigg| b(y)\bigg)
\eea
Since the remainder $b(y)$ is of degree at most $q_2-1$, we can find
the aforementioned linear combination by solving the system 
\bea
0= \L\bigg(x\le(p_n + \sum_{j=1}^{q_2} \ell_j(n) p_{n-j}\ri)\bigg| y^k \bigg) \ ,\ \ k=0,\dots, q_2-1\ .
\eea
After doing so we have that a suitable linear combination in
$x\C\{p_n,\dots,p_{n-q_1}\}$ is perpendicular to any $q=B_2 a+b$
with $\deg(a)< n-d_2-q_2$, $\deg(b)\leq q_2-1$, or -in other words -
to any $q(y)$ of degree less than $n-d_2$, thus proving the shape of the recurrence relation.

In order to clarify the genericity assumption we are imposing
we  express the above condition as a nonvanishing condition of certain
submatrices of the matrix of moments.
Indeed the polynomials $\wt p_n:= p_n + \sum_{j=1}^{q_2}\ell_j(n)p_{n-j}$ are uniquely
determined by the condition that (for $n> q_2$) 
\begin{enumerate}
\item The degree of $\wt p_n$ is $n$;
\item The polynomial $\wt p_n$ is $\L$-orthogonal to $1,y,\dots,y^{n-q_2-1}$
\item The polynomial $x\wt p_n$ is $\L$-orthogonal to $1,y,\dots, y^{q_2-1}$\ .
\end{enumerate}
This determines them (for $n > q_2$) as the following determinants (up to a nonzero multiplicative constant)
\be
\wt p_n:= c_n\det\le[
\begin{array}{cc|ccc|c}
\mu_{10} \ \dots & \mu_{1,q_2-1} & \mu_{00}& \dots & \mu_{0,n-q_2-1} &1\\
\mu_{20}\ \dots& \mu_{2,q_2-1} & \mu_{10} & \dots & \mu_{1,n-q_2-1} & x\\
\vdots &&&&& \vdots\\
\vdots &&&& & \vdots\\
\mu_{n+1,0}\ \dots &\mu_{n+1,q_2-1} & \mu_{n,0}& \dots & \mu_{n,n-q_2-1} & x^n
\end{array}
\ri]\label{223}
\ee
The genericity condition is then the nonvanishing of the principal
minor of size $n$ of the above expression, namely the nonvanishing of the determinants advocated in the statement of the theorem (eq. \ref{gen2}).

The normalization that $\wt p_n = p_n  + (lower\ degree)$  gives for
the $c_n$ of eq. (\ref{223})
\be
c_n  = \frac 1{\Delta_{n,2} \sqrt{h_n}}
\ee
Let us now check that this genericity assumption is actually
equivalent to requiring $\alpha_{d_2}(n)\neq 0\ ,\forall n$. Denoting by
$a_2,b_2$  the leading coefficients of $A_2(y), B_2(y)$ we find 
\be
b_2 \alpha_{d_2}(n) \sqrt{h_{n-d_2}} =  \L(x\wt p_n | B_2 y^{n-d_2-q_2}) =  
\L(\wt p_n|A_2 y^{n-d_2-q_2}  - \mathcal O(y^{n-d_2-1})) = \L(\wt p_n | a_2 y^{n-q_2})= \ell_{q_2}(n) a_2 \sqrt{h_{n-q_2}}
\ee 
This proves the identity (\ref{nonzero}): to prove that it does not
vanish under our genericity conditions we compute
\be
\L(\wt p_n | a_2 y^{n-q_2}) =\frac {a_2}{\Delta_{n,2} \sqrt{h_n}}   \det\le[
\begin{array}{cc|ccc}
\mu_{10} \ \dots & \mu_{1,q_2-1} & \mu_{00}& \dots & \mu_{0,n-q_2}\\
\mu_{20}\ \dots& \mu_{2,q_2-1} & \mu_{10} & \dots & \mu_{1,n-q_2} \\
\vdots &&&& \vdots\\
\vdots &&&& \vdots\\
\mu_{n+1,0}\ \dots &\mu_{n+1,q_2-1} & \mu_{n,0}& \dots & \mu_{n,n-q_2}
\end{array}
%
\ri] = \frac {a_2\Delta_{n+1,2} }{\Delta_{n,2} \sqrt{h_n}} \neq 0
\ee
{\bf Q.E.D.}\par\vskip 5pt
We can represent the previous recurrence relations in matrix form as follows
\bp
\label{multrec}
The wave vectors satisfy the following recurrence relations
\bea
x(\1 + L)\Wp = A\Wp\ ,\qquad y(\1+ M) \Ws  = B \Ws
\eea
where $L$ is the lower triangular matrix with $q_2$ subdiagonals whose matrix entries are  $L_{nm} = \ell_n(n-m)$  and $A$ is a lower Hessenberg matrix with entries $A_{nm} = \alpha_n(m-n)$ (similarly for $M, B$). The entries in the lowest and highest diagonals in $\1+L, A$ are non vanishing.
\ep
\subsection{Differential recurrence relations}
\bp
\label{pro:diffrec}
Under the genericity assumption\footnote{See Remark \ref{remark:genericity}.}  that  the principal minors of the associated moment functionals $\wc \L, \wh \L $ are all non-vanishing (or --which is the same-- the existence of  biorthogonal polynomials for $\wc \L, \wh \L $), 
the BOPs satisfy the following differential finite--term recurrence relations
\bea
&&\nabla_x \le(p_n + \sum_{1}^{q_1} \wc m_j(n+j) p_{n+j} \ri)  =
-\sum_{j=-1}^{d_1} \wc \b_j (n+j)p_{n+j}\\
&&\nabla_y \le(s_n + \sum_{1}^{q_2} \wh \ell_j(n+j)
s_{n+j} \ri)  = -\sum_{j=-1}^{d_2} \wh \a_j(n+j)s_{n+j}\\
&&\nabla_x:= \pa_x - V_1'(x)\ ,\qquad \nabla_y:= \pa_y  - V_2'(y)\ .
\eea
In matrix form we have 
\bea
\nabla_x (\1 + \wc M^t )\p = - \wc B^t \p\nonumber \\
\nabla_y (\1 + \wh L^t)\s = -\wh A^t \s\ ,\label{eq:diffrec}
\eea
where the matrices above are defined by 
\bea
\wc M_{nk} = \wc m_{n-k}(n)\ ,\qquad
\wc B_{nk} = \wc \b_{n-k}(n)\cr
\wh L_{nk} = \wh \ell_{n-k}(n)\ ,\qquad
\wh A_{nk} = \wh \a_{n-k}(n)\ . 
\eea
Note that they have the same shape as $M,B,L,A$ respectively (whence the mnemonics of the symbols).
\ep
{\bf Proof.}
We prove only the first of the two relations, the other being proved analogously.
Consider the unique (generically existing) vector $\tilde p_n$ in $\C[p_n,\dots, p_{n+q_1}]$ which is divisible by $B_1(x)$ and "monic" w.r.t. $p_{n}$ in the sense that $\tilde p_n = p_{n} + \C[p_{n+1},\dots, p_{n+q_1}]$. Writing then $\tilde p_n = B_1 q_n$ we find
\bea
{\rm e}^{V_1} \pa_x \tilde p_n {\rm e}^{-V_1} = B_1' q_n + B_1'q_n' - V'_1 B_1 q_n = B_1 q_n' - A_1 q_n.
\eea
This implies that $(-\pa_x + V_1') \tilde p_n$ is a {\em polynomial} of degree $n+d_1$ in spite of the fact that $V_1'$ is rational. Moreover
\bea
\L\bigg( (-\pa_x + V_1') \tilde p_n \bigg| y^k  \bigg) = \L\bigg( -B_1 q_n' + A_1 q_n \bigg| y^k\bigg) = 
\L\bigg( B_1 q_n \bigg| y^{k+1}\bigg) = \L \bigg(\tilde p_n \bigg| y^{k+1}\bigg) \equiv 0  \cr
 k<n-1\label{asda}
\eea
Note that  the above relation (\ref{asda}) is implicitly an assumption on the existence of polynomials $q_n$ of exact degree $n$ which are $\wc \L$--orthogonal to all lower powers of $y$: this is equivalent to saying that there must exist the BOPs  for $\wc \L$, whence our genericity assumption in the statement of the theorem. {\bf Q.E.D.}\par \vskip 5pt
For later convenience we also remark that the genericity condition we are
invoking now is also  equivalent to requiring that the vectors (the
superscript $^{(r)}$ denoting the $r$-th derivative)
\bea
\le[p_{n}^{(r)} (x_j),\dots, p_{n+q_1-1} ^{(r)}(x_j)\ri]\ ,\\
B_1(x_j)=0,\ r=0\dots r_j\ ,\qquad 
B_1(x) = b_1\prod_{j=1}^s (x-x_j)^{r_j}
\label{gencheck}
\eea
be linearly independent: indeed 
\bea
p_n + \sum_{1}^{q_1} \wc m_j(n+j)p_{n+j} = e_n\le[
\begin{array}{ccccc}
p_n(x_1) &&&&p_{n+q_1}(x_1)\\
\vdots &&&&\vdots\\
p_n^{(r_1)}(x_1) &&&&p_{n+q_1}^{(r_1)}(x_1)\\
\hline
\vdots &&&&\vdots\\
p_n^{(r_s)}(x_s) &&&&p_{n+q_1}^{(r_s)}(x_s)\\
\hline
p_n(x) & p_{n+1}(x) & \dots && p_{n+q_1}(x)
\end{array}
\ri]
\eea
where $e_n$ is the inverse of the $(q_1+1,1)$--cofactor of the above matrix.
The proposition can be rewritten for the wave vectors as follows
\bp
\label{diffrec}
The wave vectors satisfy the following differential equations
\be
\pa_x (\1 + \wc M^t) \Wp = -\wc B^t \Wp\ ,\qquad 
\pa_y(\1 + \wh L^t) \Ws = -\wh A^t \Ws\ ,
\ee
where $\wc M_{nk} = \wc m_{k-n}(n)$ and $\wh A_{nk} = \wh \a_{k-n}(n)$
(and similar expressions for $\wh M, \wh B$).
\ep
The matrices $\wc M, \wc B, \wh L, \wh A$ play the same role of $M,B$ and $L,A$ for the moment functionals $\wc \L$ and $\wh\L$ respectively.
\bp
The vectors of polynomials \footnote{The expressions $(\1 +\wh  L)^{-1}$ etc. are defined by the geometric series; since $\wh L$ is strictly lower triangular, such geometric series is entry--wise well defined.}
\be
\wh \p(x) := (\1 + \wh L)^{-1} \p\ ,\qquad \wh \s(y) := \frac 1{B_2(y)} (\1 + \wh L^t) \s(y)
\ee
(where $\wh L$ (and $\wc M$) are defined by eqs.(\ref{eq:diffrec}) of
Prop. \ref{pro:diffrec}) 
are the biorthogonal polynomials for $\wh \L$. 
Similarly the vectors of polynomials 
\be
\wc \p(x):= \frac 1 {B_1(x)} (\1 + \wc M^t) \p\ ,\qquad 
\wc \s(y):= (\1 + \wc M)^{-1} \s
\ee
are the biorthogonal polynomials for $\wc \L$
\ep
{\bf Proof.}
The two statements are completely parallel and hence we prove only the first.

By definition of the matrix $\wh L$ in Prop. \ref{pro:diffrec} the polynomial entries of $(\1 + \wh L^t)\s$ are all divisible by $B_2$, therefore $\wh \s$ is indeed a vector of polynomials.
Next we have (using an obvious matrix notation)
\be
\wh \L \le(\wh \p \bigg | \wh \s ^t\ri) = \L \le((\1 + \wh L)^{-1}\p
\bigg |\s^t (\1 + \wh L)\ri) =(\1 + \wh L)^{-1} \L \le(\p \bigg
|\s^t\ri) (\1 + \wh L) = \1\ \hbox{ {\bf Q.E.D}}
\ee
We also have
\bl
The matrices $L,A,M,B$ and the matrices $\wh L, \wh A, \wc M, \wc B$ are related by 
\be
A (\1 + \wh L) = (\1 + L)\wh A\ ,\qquad
B(\1 + \wc M) = (\1 + M)\wc B \ .
\ee
\el
{\bf Proof.}
Once more we prove only the first.
\bea
A (\1 + \wh L) =\L\le( A \p   \bigg | \s^t(\1 + \wh L)      \ri) = \L\le(  x(\1 + L)\p   \bigg | \s^t(\1 + \wh L)        \ri) =\cr
=  \L\le(  x(\1 + L)\p   \bigg | B_2\wh\s^t     \ri) =\L\le(  (\1 + L)\p   \bigg | (-B_2\pa_y +A_2)\wh\s^t     \ri) =
\L\le(  (\1 + L)\p   \bigg | -\nabla_yB_2\wh\s^t     \ri) =\cr
=\L\le(  (\1 + L)\p   \bigg |-\nabla_y \s^t(\1 + \wh L)     \ri)
=\L\le(  (\1 + L)\p   \bigg |\s^t A     \ri) = (\1 + L) \wh A \qquad
\hbox{{\bf Q.E.D.}}
\eea
\bl
\label{assorec}
The associated wave vectors $\wh \p,\wh \s$ and $\wc \p,\wc \s$ satisfy
\bea
x(\1 + \wh L)\wh \p = \wh A \wh \p\cr
y(\1 + \wc M)\wc \s = \wc B \wc \s
\eea
Moreover, under the same genericity assumptions
 \be
 \wc m_{q_1}(n)\neq 0 \neq \wh \ell_{q_2}(n) \qquad \forall n
 \ee
\be
b_2\wh \alpha_{d_2}(n)\sqrt{\wh h_{n-d_2} }= a_2 \wh \ell_{q_2}(n) \sqrt{\wh h_{n-q_2}}
\neq 0\label{nonzerodual}
\ee
\el
{\bf Proof.}
Recalling that $\wh \p = (\1 + \wh L)^{-1} \p$ (by definition), we find 
\bea
x(\1 + \wh L)\wh \p = x \p = (\1 + L)^{-1} A \p = \wh A (\1 + \wh L)^{-1} \p = \wh A \wh \p\ .
\eea
The relations (\ref{nonzerodual}) for the moment functionals $\wh \L, \wc \L$ 
are proved in exactly the same way relations (\ref{nonzero}) are
proved for $\L$. {\bf Q.E.D.}\par \vskip 5pt 
We can summarize all the relations collected so far in Table \ref{table}.
\begin{table}
\begin{tabular}{c|c|c|c}
Functional & BOPs & Mult. rec. & Diff. Rec. \\
$\ds \L(\bullet |\bullet) $ &$\ds  \p,\s$  & $\ds \begin{array}{c}
x(\1 + L)\p  = A\p\\
y(\1 + M)\s = B\s
\end{array} $ & $\ds \begin{array}{c}
\nabla_x (\1 + \wc M^t)\p  =-\wc B^t\p\\
\nabla_y (\1 + \wh L^t)\s = -\wh A^t\s
\end{array} $ \\
\hline 
$\ds \wh \L (\bullet| \bullet) =  \L(\bullet |B_2\bullet) $ & $\ds 
\begin{array}{c}
\wh \p := (\1 + \wh L)^{-1} \p\\
\wh \s := \frac 1{B_2} (\1 + \wh L^t)\s
\end{array}
$ &  $\ds \begin{array}{c}
x(\1 + \wh L)\wh \p  = \wh A\wh \p\\
\star
\end{array} $ &
$\ds \begin{array}{c}
\nabla_x (\1 + \cs M^t )\wh \p  = -\cs B^t \wh \p\\
\star
\end{array} $
\\
\hline 
$\ds \wc \L (\bullet | \bullet) =  \L(B_1\bullet |\bullet) $ & $\ds
\begin{array}{c}
\wc \p = \frac 1 {B_1} (\1 + \wc M^t)\p\\
\wc \s = (\1 + \wc M)^{-1} \s
\end{array}$ &  $\ds \begin{array}{c}
\star\\
y(\1 + \wc M)\wc \s  = \wc B\wc\s
\end{array} $ &
 $\ds \begin{array}{c}
\star\\
\nabla_y (\1 + \cs L^t)\wc \s  = -\cs A^t\wc\s
\end{array} $ \\
\hline 
$\ds \cs \L (\bullet | \bullet) =  \L(B_1\bullet |B_2 \bullet) $ & $\ds
\begin{array}{c}
\cs \p =\le\{
\begin{array}{c}
 \frac 1 {B_1} (\1 + \wt M^t)\wh \p\\
 (\1 + \cs L)^{-1} \wc \p
 \end{array}\ri.\\
\cs\s =
\le\{
\begin{array}{c}
 (\1 + \wt M)^{-1}\wh \s\\
 \frac 1{B_2} (\1 + \wt L^t)\wc \s
 \end{array}\ri.
\end{array}$ &  $\ds \begin{array}{c}
x(\1 + \cs L)\cs \p = \cs A \cs \p\\
y(\1 + \cs M)\cs \s  = \cs B\cs\s
\end{array} $ &
 $\star $ 
\end{tabular}
\bea
(\1 + L) \wh A = A (\1 + \wh L)\qquad
(\1 + \wh L) \cs A = \wh A (\1 + \cs L)\\
(\1 + M)\wc B = B (\1 + \wc M)\qquad
(\1 + \wc M)\cs B = \wc B (\1 + \cs M)
\eea
\caption{ the various recurrence relations.}
 \label{table}
\end{table}
Here the matrices $A, \wh A, \cs A$ are lower--Hessenberg matrices with $d_2$ nontrivial sub-diagonals, $B,\wh B, \cs B$ are lower--Hessenberg with $d_1$ nontrivial sub--diagonals. The matrices $L,\wh L, \cs L$ and $M, \wh M, \cs M$ are strictly lower triangular matrices with $q_2$ or $q_1$  nontrivial subdiagonals respectively.

The $\star$'s mean that there are (possibly under similar genericity requirements for the corresponding functional) similar relations as in the corresponding box on the first line, for which however we do not need to define symbols for our purposes.
\br
\label{remark:genericity}
We now address the genericity assumptions invoked in Thm. \ref{thmrecmul} and Prop. \ref{pro:diffrec} in the case of real potentials and support on the real axes as discussed in the Introduction.
For Prop. \ref{pro:diffrec} the assumption is simply the existence of the BOPs for the associated moment--functional $\wh L$; in the case of real potentials with supports on the real line one can follow \cite{ercoken} and show that BOPs do exist. Since in $\L$ and $\wh \L$  have the same supports and $B_2(y)$ would be positive on the support (provided that none of the higher-multiplicity zeroes lie  within the support) then one can conclude that for the case of relevance to the Hermitean two--matrix model the Prop. \ref{pro:diffrec}) is always valid.

Less transparent is the extent of the limitation imposed by the genericity assumption (\ref{gen2})  used in Thm. \ref{thmrecmul}; the polynomials $\wt p_n$ appearing in the proof of said theorem play the same r\^ole in respect to $p_n$'s as the $p_n$'s play regarding the $\wh p_n$'s (see Table \ref{table}); this means that they could be obtained from a functional $\L'(\bullet | \bullet) = \L(\bullet | B_2^{-1} \bullet)$.

Note, however, that since the matrices $L$ and $A$ are not uniquely defined in the $d_2\times d_2$ principal minor (in the Theorem we fixed the ambiguity by setting the corresponding block of $L$ to zero), the above possibility is not the sole choice.

Moreover, if some of the hard-edges zeroes of $B_2(y)$ belong to the real axis (i.e. if there are hard-edges on $\R$) then such choice is not viable because the integral defining $\L'$ would be divergent at the hard-edges. In this case this simply means that $\wt p_n$'s do not belong to a biorthogonal pair for some semiclassical functional but are just defined (for $n > q_2$)  by eq. (\ref{223}) and the genericity issue cannot be resolved easily.

Vice-versa, in the case none of the hard-edge zeroes of $B_2$ belong to $\R$ then $\L'$ indeed exists and is a semiclassical functional (with $V_2$ replaced by $V_2+\ln B_2$). The existence of the corresponding BOPs (hence the verification of the genericity assumption) then follows again from the result in  \cite{ercoken}. 
\er
%
\section{Adjoint differential equations and the bilinear concomitant} 
\label{bilconco}
In this section we recall some results  which
--although simple-- I was not able to find in the literature.
We consider a $n$th order differential equations of the form 
\be
\bigg(A(\pa_x) - x B(\pa_x)\bigg)f(x) = 0\label{ODE}
\ee
where $A(D)$ and $B(D)$ are polynomials and $n=
\max(\deg(A),\deg(B))$: the reader should keep in mind
the polynomials $A_i,B_i$ of our matrix model.
If we look for solutions written as  ``Fourier--Laplace''
transforms
\be
f_\Gamma(x):= \int_\Gamma {\rm d}y\, {\rm e}^{xy-V(y)}\ ,
\label{solODE}
\ee
--where the contour of integration is so far unspecified--, formal manipulations
involving integration by parts show that 
\be
V'(y) = \frac {A(y) + B'(y)}{B(y)}\ .
\ee
We point out that the relation between $V$ and $A,B$ in these formul\ae\ is exactly the same as the relations between the $V_i$'s and $A_i,B_i$'s  of the first part of the paper.

In the situation of interest to us we will have $A,B$ {\bf reduced}
\bd
Two polynomials $A,B$ are called {\bf reduced} if the only zeroes that they share (if any) are amongst the simple zeroes of $B$.
\ed
\bl
\label{lemmadual}
Two polynomials $A,B$ are reduced if and only if $A \pm B'$ and $B$ are.
\el
{\bf Proof}
Suppose $A, B$ are  reduced. 
If $c$ is a common zero of $A$ and $B$ (hence simple for $B$) then  $B' (c)\neq 0$:
therefore, $A(c)\pm B'(c)\neq 0 $  (because $A(c)=0$ and $B'(c)\neq 0$).
So $A\pm B'$ and $B$ do not share this particular zero.

Now let $\tilde c$ be a common zero of $A \pm B'$ and $B$: if it were not a simple zero of $B$ then $B'(\tilde c)=0 $ and hence also $A(\wt c )=0$. But this contradicts that  $A$ and $B$ are reduced because they share a zero which is non simple for $B$.

Viceversa: suppose $\tilde A:= A \pm  B'$ and $B$ are reduced. Then by the above
$\tilde A \mp B' = A$  and B are reduced. Q.E.D.\par \vskip 5pt

This "duality" of the notion of reducedness will be important when considering the adjoint differential operator.

We now remark that $V'$
is a rational function with poles at a {\bf subset} of the zeroes of
$B$
\bea
B(y) = c\prod_{j=1}^{r} (y-b_j)^{m_j+1} \ ,c\neq 0\ ,  \deg B =
\sum_{j=1}^r m_j\ ,\ \ m_j\in \N\ .\\
V'(y) = \sum_{\ell=0}^{d} v_{\ell+1} y^\ell  - \sum_{j\in
  J\subseteq {1,\dots,r}}
\sum_{k=0}^{m_j} \frac {t_{k,j}}{(y-b_j)^{k+1}}\\
{\rm e}^{-V(y)} = \prod_{j\in J} (y-b_j)^{t_{0,j}} \exp\bigg[
\sum_{\ell=0}^{d} \frac { v_{\ell+1}}{\ell+1}  y^{\ell+1} + \sum_{j\in J}
 \sum_{k=1}^{m_j} \frac {t_{k,j}}{k(y-b_j)^{k}}
\bigg] \\
W(y):= {\rm e}^{-V(y)}\\
d:= \deg (A)-\deg(B)\ .
\eea
[Here it is understood that if $\deg(A)<\deg(B)$ then the first sum in $V'$  is absent.]
 
Some of the zeroes of $B(y)$ may appear also as zeroes of
$A(y)+B'(y)$ and hence in the partial fraction expansion of $V'$ those
points do not appear. Since $A,B$ are reduced, all multiple
zeroes of $B$ are not shared with $A+B'$. We will call the zeroes of
$B$ which are common with $A+B'$ the {\bf hard-edge points} (note that
not all simple zeroes of $B$ are hard-edge points, but all hard-edge
points are simple zeroes). 

We now define some sectors $S_k^{(j)}$, $j=1,\dots p_1$, $k=0,\dots m_j-1$.
around the multiple zeroes of $B$ ($b_j$ for which $m_j>0$) 
in such a way that 
\be
\Re\le(V(y)\ri) \mathop{\longrightarrow}_{
\shortstack{{\scriptsize $y\to b_j,$}\\
{\scriptsize  $y\in S
^{(j)}_k$}}}
+\infty\ .
\ee
The number of sectors for each pole is the degree of that pole in the
exponential part of $W(x)$, that
is $d+1$ for the pole at infinity and $g_j$ for the $j$-th pole.
Explicitly 
\bea
&&\hspace{-0.5cm}S^{(0)}_k := \le\{ y:\in \C;\ \ \frac {2k\pi -\frac \pi
2+\epsilon}{d+1}< \arg(y)+\frac{\arg(v_{d+1})}{d+1} < \frac {2k\pi +\frac \pi
2-\epsilon}{d+1} \ri\} ,\cr
 &&\ k=0\dots d\ ;\nonumber \\
&&\hspace{-0.5cm}S^{(j)}_k := \le\{ y:\in \C;\ \ \frac {2k\pi -\frac \pi
2+\epsilon}{m_j}< \arg(y-b_j)+\frac{\arg(t_{m_j,j})} {m_j} < \frac {2k\pi +\frac \pi
2-\epsilon}{m_j} \ri\},\label{sectinf}\\
&& k=0,\dots,m_j-1,\ \ j\in J \ .\nonumber
\eea
These sectors are defined precisely in such a way that approaching any
of the
essential singularities (i.e. an $b_j$ such that $m_j>0$) the function
$W(y)$ tends to zero faster than any power.
\subsection{Definition of the contours}
\label{defcont}
The contours we are going to define are precisely the type of contours
$\Gamma_{x,\mu}, \Gamma_{y,\nu}$ entering the definition of the
bimoment functional $\L$.
Let $A,B$ be reduced: we then define $n=\max(\deg (A),\deg (B))$ contours.
The definition of the contours follows directly \cite{jat, shapiromiller}. 
We first remark that the weight $W(y)$ is --in general-- multi-valued since it
contains powers like $(y-c)^t$ with non-integer $t$; the
multivaluedness is multiplicative and in fact is not very important
which branch one chooses in the definition of the integrals  
(\ref{solODE}) since different choices correspond to multiplying the same function by a nonzero constant.
Nonetheless it will be convenient at some point to have a reference normalization for the integrals and hence 
we define some cuts so as to have a simply connected domain where $W(y)$ is single-valued.
We do so by removing semi-infinite arcs extending from each branch-point
of $W(y)$ to infinity: for convenience we choose the cuts approaching
each singularity in one of the sectors, for example $S_0^{(j)}$, and
approaching infinity within $S_0^{(0)}$.  
If $\deg(A)\leq \deg(B)-1$ then no sector is defined at $\infty$ and
then we just choose arbitrarily an asymptotic direction for these
cuts. 
Note that if $\deg(A)\leq \deg (B)-2$ then the sum of the finite
residues of $V'{\rm d}y$ is zero, hence we could define the cuts as
finite arcs joining in a chain the finite branch-points of $W(y)$: the
resulting domain is not simply connected, however $W(y)$ is single
valued in such domain precisely because of  the vanishing of the sum
of the residues of its logarithmic derivative. 
We will denote by $\mathcal D$ the  connected domain obtained
after such surgery.

In the following our primary focus is on the case $\deg (A)\geq \deg
(B)+1$ and we leave to the reader to check the literature
\cite{shapiromiller} for the remaining cases (only minor modifications
are needed).

\begin{figure}
\epsfxsize 12cm
\epsfysize 12cm
\epsffile{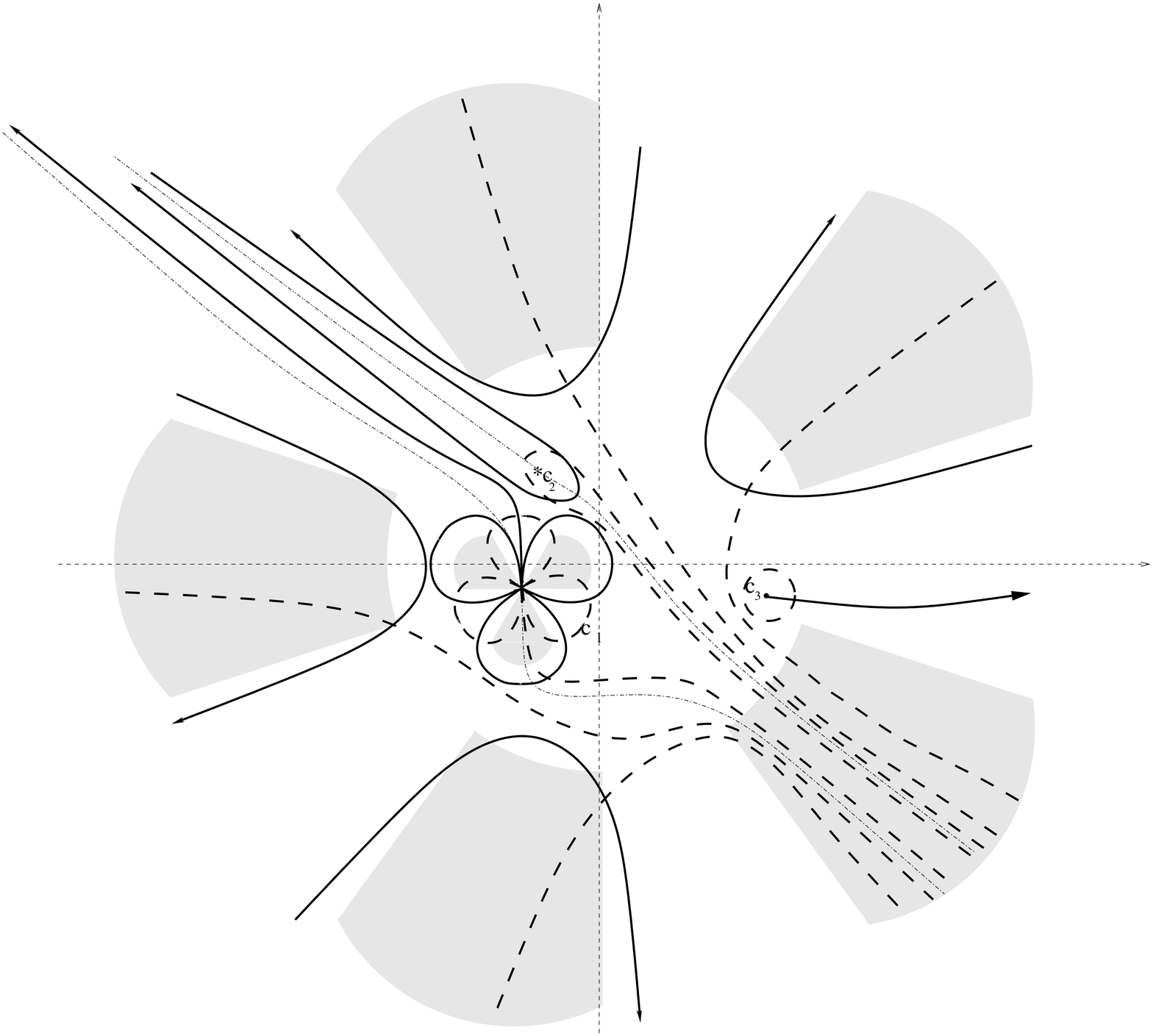}  
\caption{\label{contours} 
\fontfamily{cmss}
\fontsize{11pt}{15pt}
\selectfont
An example of contours $\Gamma$ and $\wh
  \Gamma $ for a pair of reduced adjoint differential operators. The
  thick contours are the admissible ones for $L$ while the thick dashed ones
  are the admissible ones for $L^\star$. Also shown in the picture are
  the cuts for $W(y)$ and $\wh W(s)$ (line-dotted thin lines).} 
  \label{figone}
\end{figure}

\begin{enumerate}
\item For any zero $b_j$ of $B$ for which there is {\em no essential singularity} in $W$
we have two cases
\begin{enumerate}
\item If  $b_j$ is a  branch point (i.e. $t_{0,j}\in \C\setminus \Z$)
we take a loop (referred to as a {\bf lasso}) starting at infinity in some fixed sector
(e.g. $ S_{0}^{(0)}$)  encircling the singularity and going back to infinity in the same
sector.
\item If $b_j$ is a pole of $W$ (i.e. $t_{0,j}\in \{-1,-2,-3,\dots\} $) then we take a small circle around it.
\item If  $b_j$ is a  regular point ( namely $t_{0,j}\in \{0,1,2,\dots\}$) we
take a line joining
$b_j$ to infinity and approaching $\infty$ in the same sector $S_{0}^{(0)}$
 as before (this case includes the hard-edge points for which we may say that $t_{0,j}=0$).
\end{enumerate}
\item For any  multiple zero $b_j$ for which there is an essential singularity
(i.e. for which $m_j>0$) we define $m_j$ contours (which we call the {\bf petals}) starting from $b_j$
in the sector $S_0^{(j)}$ 
and returning to $b_j$ in the next (counterclockwise) sector. Finally
we join the singularity $b_j$ to $\infty$ by a path (called the {\bf stem}) approaching $\infty$
within the sector $ S^{(0)}_{0}$ chosen at point 1(a).
\item If $\deg(A)\geq \deg(B)+1 $  we define  $b_0:=\infty$ and  we take $d:= \deg (A)-\deg (B)$ contours starting at $X_0$ in the
sector $S^{(0)}_k$ and returning at $X_0$ in the sector
$S^{(0)}_{k+1}$.
\end{enumerate}
The reasons for the "floral" names should be clear by looking at an example like the one in Fig. \ref{contours}. 
Cauchy's theorem grants us large freedom in the choices of such contours; we use this freedom so that the contours do not intersect each other in  $\C\setminus\{b_j\}_{j=1,\dots,\deg(B)}$ and do not cross the chosen cuts.

We will refer to these contours collectively as {\bf admissible contours} for the differential $W(y){\rm d}y$. 
Note that we have defined exactly $n=\max(\deg(A),\deg(B))$ contours.

It is a straightforward check to see that 
\be
f_\Gamma(x):= \int_\Gamma {\rm d}y \,{\rm e}^{xy-V(y)} = \int_\Gamma {\rm e}^{xy} W(y){\rm d}y\ ,
\ee
all satisfy the differential equation (\ref{ODE}): in these checks one
is always allowed to perform integration by parts discarding all
boundary terms because of the properties of the contours. We leave
this check to the reader. 

The content of \cite{shapiromiller} (and of the fix contained in
\cite{jat}) was to show that these functions are also linearly
independent, hence providing a basis for the solution space. 
\subsection{Adjoint differential operators and the bilinear concomitant}
In general, given a $n$-th order linear operator with polynomial coefficients
\be
L:= \sum^n_j a_j(x) \pa_x^j\ ,
\ee
its classical adjoint is defined as 
\be
L^\star := \sum_j^n (-\pa_x)^j a_j(x)\ .
\ee
Between the solution spaces of a pair of adjoint such operators
Legendre defined a nondegenerate pairing called the {\bf bilinear
  concomitant} \cite{ince}. We will show that this pairing for our class of
reduced operators admits a natural interpretation as
intersection pairing. 

We begin by noticing that in our case the pair of adjoint operators is written
\be
L:= A(\pa_x) - x B(\pa_x)\ ,\qquad L^\star := A(-\pa_x) - B(-\pa_x)x\ .
\ee 
Since $A,B$ are reduced then $L^\star$ is also reduced since 
\be
L^\star = A(-\pa_x) - B'(-\pa_x) -x B(-\pa_x)
\ee
 in view of Lemma \ref{lemmadual} (here the polynomials are
 $A(-y)-B'(-y)$ and $B(-y)$ which are clearly reduced iff $A(z)-B'(z)$
 and $B(z)$ are). Therefore $L^\star$ is in the same class of
 operators as $L$ and can be solved by contour integrals in the same
 way. 
The solutions of $L^\star g=0$ are of the form 
\bea
g =\int_{\wh \Gamma} {\rm e}^{-xs+\hat V(s)}{\rm d}s\\
\wh V(s)':= \frac {A(s)}{B(s)} = V'(s) - (\ln B(s))'\ .
\eea
An inspection shows that the sectors around the multiple zeroes
of $B(s)$ where $\Re(\hat V(s))\to -\infty$ are precisely the
complementary sectors defined in (\ref{sectinf}) for $V$. 
We normalize $\wh V(s)$ by choosing the integration constant in such a way that 
\be
\wh W(s):= {\rm e}^{\wh V(s)}  = \frac 1{B(s)} {\rm e}^{V(s)}
\label{dualweight}
\ee
(here ${\rm e}^{V}$ is supposed to be defined on the simply connected domain $\wh {\mathcal D}$).
One then proceeds in the definition of the admissible contours $\wh
\Gamma$ for the weight $\wh W(s)$  and of the simply connected domain
$\wh {\mathcal D}$  in exactly the same way used for $W(y)$. We make
the following important remarks:
\begin{enumerate}
\item If $b_j$ is a hard-edge point for $W(y)$ (i.e. it is a zero of
  $B(y)$ but a regular point for $W(y)$ where $W$ does not vanish)
  then $b_j$ is a simple pole of $\wh W(s)$.  
\item If $b_j$ is a zero of multiplicity $m$ of $W(y)$ (i.e. a simple
  zero of $B(y)$ such that the residue of $(A+B')/B{\rm d}y$ is  a
  negative integer) then it is a pole of order $m+1$ for $\wh W(s)$. 
\item In all other cases, the type of singularity of $W$ and $\wh W$ is the same (logarithmic branch-points or essential singularities of the same exponential type).
\item The intersection $\mathcal D\cap \wh{\mathcal D}$ is the
  disjoint union of simply connected domains where $W(y)\wh W(y)B(y)$
  is constant. These constants depend only on the residues of
  $V'(y){\rm d}y$ mod $\Z$.  
\end{enumerate}
These observations and the fact that $B(y) W(y)\wh W(y)$ is locally constant (where they are both defined) follows immediately from their definition and eq. (\ref{dualweight}).

From the definitions of the contours it is not difficult to realize that dual contours can be chosen such that 
\begin{enumerate}
\item For each flower (petal + stem) one can choose a dual flower
  whose elements intersect only the arcs of the given flower. (This
  includes the petals at $\infty$, in the case $\deg(A)\geq
  \deg(B)+1$). 
\item For each pole $c$ of $W(y)$ (whose corresponding admissible
  contours $\Gamma$ is a small circle) the dual admissible  contour
  for $\wh W(s)$ is  a semi-infinite arc starting at $c$ and going to
  $\infty$ and can be chosen so that it intersects only its dual. 
\item For each zero or hard-edge point $a$ of $W(y)$ (whose
  corresponding admissible contour is a semi-infinite arc starting at
  $a$) the dual admissible contours for $\wh W(s)$ (which is a small
  circle around $a$) intersects only $\Gamma$. 
\item For each non-essential other singularity of $W(y)$ (i.e. a
  simple zero $c$ of $B(y)$ such that the residue of $(A+B') /B{\rm
    d}y $ is in $\C\setminus \Z$), where the admissible contour
  $\Gamma$ is a lasso around $c$, the dual loop $\wh \Gamma$ (also a
  lasso around $c$) is also chosen so that it intersects only the dual
  lasso (at {\em two} points). 
\end{enumerate} 
\bl
Consider the two adjoint differential equations
\bea
\bigg(A(\pa_x) -xB(\pa_x)\bigg)f(x)=0\\
\bigg(A(-\pa_x)- B(-\pa_x)x\bigg) g(x) = 0\ .
\eea
The solutions are of the form\footnote{The formula depends on the integration constant in $V$, namely these solutions are defined up to multiplicative constants since they are solutions of a homogeneous linear ODE.}
\bea
&& f(x)= f_\Gamma (x):= \int_\Gamma {\rm e}^{- V(y)+xy}{\rm d}y\ ,\ \ \  V:= \int \frac
{A(y) +B'(y) }{B(y)}{\rm  d}y\label{solf}\\
&& g(x) = g_{\wh \Gamma}(x) := \int_{\wh \Gamma} {\rm e}^{\wh V(s)-xs}{\rm }ds \ ,\ \ \
 \wh V(s):= \int \frac {A(s)}{B(s)}{\rm d}s\label{solg}
\eea
Then the following expression is constant and defines a nondegenerate
bilinear pairing (the {\bf bilinear concomitant}) between the
solutions spaces of the two adjoint equations:
\bea
\mathcal B(f,g):= \int_{\wh \Gamma}\int_\Gamma
\le[\bigg(B(y)-B(s)\bigg)  \le[ \frac x{y-s} - \frac 1 {(y-s)^2} \ri] -
\frac {A(y)-A(s)-B'(s)}{y-s}\ri] {\rm e}^{x(y-s) -  V(y)+\wh V(s)} {\rm d}y{\rm d}s
\eea
\el
{\bf Proof.}
The integral representation of the solution is easily verified. We now
write
\bea
&&0\equiv g(x)\int_{ \Gamma} \le(xB(y)-A(y)\ri){\rm e}^{- V(y)+xy}{\rm d}y\\
&&0\equiv f(x)\int_{\wh \Gamma} \le(xB(s)-A(s)-B'(s) \ri){\rm
  e}^{ \wh V(s)-xs}{\rm d}s
\eea
We take the difference and obtain
\bea
0\equiv \int_{\wh \Gamma}
\int_{\Gamma}\le(x(B(y)-B(s))-(A(y)- B'(s)-A(s))\ri)  {\rm e}^{x(y-s) -
  V(y)+\wh V(s)}{\rm d}y {\rm d}s
\eea
It is promptly seen that the integrand of this double integral  is
absolutely summable w.r.t. the arc-length parameters along $\Gamma$ and
$\wh \Gamma$, hence we can integrate w.r.t. $x$ under the integral
sign, thus obtaining the bilinear concomitant; 
\bea
\int_{\Gamma} \int_{\wh \Gamma}\le((B(y)-B(s))\le(\frac x{y-s} -
\frac 1{(y-s)^2}\ri)-\frac{A(y) - B'(s)-A(s)}{y-s}\ri)  {\rm e}^{x(y-s)
  - V(y)+\wh V(s)}{\rm d}s{\rm d}y
\eea
Note that the expression under integration is regular at $y=s$, and is
--in fact-- a {\bf polynomial} in $y,s$
\bea
\le((B(y)-B(s))\le(\frac x{y-s} -
\frac 1{(y-s)^2}\ri)-\frac{A(y) - B'(s)-A(s)}{y-s}\ri)  \m\sim_{y\to s}
xB'(s) - \frac 1 2 B''(s) - A'(s) + \mathcal O(y-s)\nonumber
\eea
In particular the integrand is absolutely integrable w.r.t. the
arc-length parameters and hence the order of integrations is
irrelevant. 
This concludes the proof. {\bf Q.E.D.}\par \vskip 5pt

The bilinear concomitant is --in a certain sense-- an integral representation of the
intersection pairing of the contours of integration.
To make this statement more precise we first prove the following standard
\bl
\label{interlemma}
Let $\Omega(y,s)$ be a meromorphic  function  $\mathcal D\times \wh
{\mathcal D}$ where ${\mathcal D}$ and $\wh {\mathcal D}$ are simply
connected domains and with the only singularities  being a double pole as $y\to s$ (in $\mathcal D\cap \wh {\mathcal D}$). Suppose that in each connected component of
$\mathcal D\cap \wh {\mathcal D}$ there is a constant $c$ such that  
\be
\Omega(y,s) = \frac {c}{(y-s)^2} + \mathcal O(1)
\ee
as $y\to s$ within the intersection domain.
 Let $\Gamma\subset \mathcal D$ be a smooth curve  such that 
\be
\int_\Gamma \Omega(y,s){\rm d}y \equiv 0
\ee 
Let $\wh \Gamma \subset \wh {\mathcal D}$ be a curve of finite length
intersecting once $\Gamma$ at $p$  and oriented positively
w.r.t. $\Gamma$: then  
\be
\label{terse}\int_{ \Gamma} {\rm d}y \int_{\wh \Gamma} {\rm d}s \Omega(y,s) = 2i\pi c(p)
\ee
\el
{\bf Proof.}
The integral 
\be
f(s):=\int_\Gamma \Omega{\rm d}y
\ee
defines --in principle-- different holomorphic  functions in the
connected components of $\wh {\mathcal D} \setminus \Gamma$: the
difference among them -however- is the residue  
\be
\res{y=s}\Omega(y,s){\rm d}y 
\ee 
which is zero by the assumption on $\Omega$. Hence the analytic
continuations of $f(s)$ from one component to the other all
coincide. In our case they are all zero. The key fact is that, since
$\Omega$ is singular on the diagonal, the orders of integration
matters (otherwise (\ref{terse}) would give zero by interchanging the
order of integration).

We compute the integral as a limit of regular integrals where we can interchange the order of integration
\be
(\ref{terse}) = \lim_{\epsilon\to 0} \int_{ \Gamma_\epsilon} {\rm d}y \int_{\wh \Gamma} {\rm d}s \Omega(y,s) \ ,
\ee
where $\Gamma_\epsilon$ is the curve (or union of curves) obtained by
removing a small $\epsilon$-arc (which we denote by $\Gamma^\epsilon$,
i.e. an arc from $p-\epsilon$ to $p+\epsilon$, where these two points lie on the curve $\Gamma$ at distance $|\epsilon|$ from the intersection and the direction of $\epsilon$ is the same as the orientation of $\Gamma$) around the
intersection point $p$. This allows us to interchange the order of
integration under the limit sign 
\bea
&\&  \lim_{\epsilon\to 0} \int_{ \Gamma_\epsilon} {\rm d}y \int_{\wh \Gamma} {\rm d}s \Omega(y,s) = 
  \lim_{\epsilon\to 0}  \int_{\wh \Gamma} {\rm d}s \int_{ \Gamma_\epsilon} {\rm d}y \Omega(y,s)  = 
-  \lim_{\epsilon\to 0}  \int_{\wh \Gamma} {\rm d}s \int_{ \Gamma^\epsilon} {\rm d}y \Omega(y,s)  = \cr
&\& =
-\lim_{\epsilon\to 0} \int_{\wh \Gamma} {\rm d}s \int_{
  \Gamma^\epsilon} {\rm d}y \le(\frac{c(p)}{(y-s)^2} + \mathcal
O(1)\ri) =  
-\lim_{\epsilon\to 0} \int_{\wh \Gamma} {\rm d}s \int_{ \Gamma^\epsilon} {\rm d}y \frac{c(p)}{(y-s)^2} 
\eea
where we have dropped the $\mathcal O(1)$ part since the length of
$\wh \Gamma$ is finite and that of $\Gamma^\epsilon$ tends to zero.
In the last expression the inner integral is --strictly speaking--
defined only for $s\neq p$: however on the "left" and "right" the
result is the same and gives 
\bea
&\&-\lim_{\epsilon\to 0} \int_{\wh \Gamma} {\rm d}s \int_{p-\epsilon}^{p+\epsilon } {\rm d}y \frac{c(p)}{(y-s)^2}  = 
c(p) \lim_{\epsilon\to 0} \int_{\wh \Gamma} {\rm d}s \le(\frac 1{b-p-\epsilon} - \frac 1{b-p+\epsilon}\ri)  = \cr
&\&=
c(p) \lim_{\epsilon\to 0} \ln\le(\frac {b-p-\epsilon}{a-p-\epsilon} \ri) - \ln \le(\frac {b-p+\epsilon}{a-p+\epsilon} \ri)
\eea
In this last limit the logarithms appearing have different branches:
in particular the second differ by $2i\pi$ from the first, hence the
result follows by taking the limit. {\bf Q.E.D.}\par
\begin{figure}
\epsfxsize 5cm
\epsfysize 5cm 
\epsffile{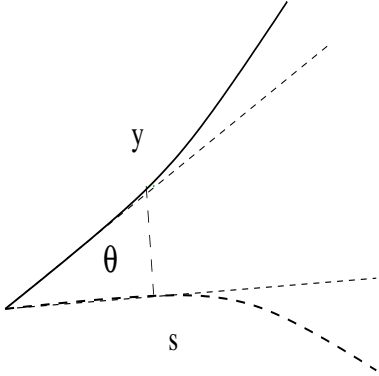}
\caption{Illustration for estimate (\ref{3-41})}
\label{fig2a}
\end{figure}

\vskip 5pt 
We now come back to the computation of the concomitant: first of all,
since we know that the result is independent of $x$ we set $x=0$, so that we have to compute 
\be
\mathcal B(f,g):= \int_{\wh \Gamma}\int_\Gamma   \le[- \frac {B(y)-B(s)} {(y-s)^2}-
\frac {A(y)-A(s)-B'(s)}{y-s}\ri] {\rm e}^{-  V(y)+\wh V(s)} {\rm d}y{\rm d}s
\ee
We have already remarked that this integral can be computed in either
orders and gives the same result. We express it in terms of  
\bea
\mathcal B(f,g)  &\& = (2) - (1)\\
&\& (1):=  \int_\Gamma {\rm d}y \int_{\wh \Gamma}{\rm d}s    \le[ \frac {B(y)} {(y-s)^2}-
\frac {A(y)}{y-s}\ri] W(y)\wh W(s)\\
&\& (2):=  \int_\Gamma {\rm d}y \int_{\wh \Gamma}{\rm d}s  \le[ \frac {B(s)} {(y-s)^2}-
\frac {A(s)+B'(s)}{y-s}\ri] W(y)\wh W(s) \eea
The integral $(2)$ is zero because the inner integral w.r.t. $s$
defines (for $y\not \in \wh \Gamma$) the identically zero function, as
it is easily seen after an integration by parts. 
The integral $(1)$ is computed using Lemma \ref{interlemma} after noticing that 
\be
\Omega(y,s) :=   \le[ \frac {B(y)} {(y-s)^2}-
\frac {A(y) }{y-s}\ri] W(y)\wh W(s)  = \frac {B(s) W(s)\wh W(s)}{(y-s)^2} + \mathcal O(1)\ .
\ee
and hence satisfies the condition of the Lemma for $\Omega$. The
contour $\Gamma$ satisfies the condition of the Lemma. The contour
$\wh \Gamma$ is not necessarily of finite length, but we can take only
a small arc around the point of intersection and the remainder  will
be computed to be zero by interchanging the order of the integrals. 
To rigor one should also consider the common endpoints of contours
like the petals and dual petals: it is easily seen, however that those
points do not correspond to a singularity of the integrals (w.r.t. the
arclength parameters) because of the fast decay of the weights $W$ and
$\wh W$. For example, if the two contours $\Gamma, \wh\Gamma$ form an
angle $\theta\in [0+\epsilon,\pi-\epsilon]$ (asymptotically) near a point $b$
(where $W,\wh W$ have an essential singularity) then \\ 
\be
\le|\frac {W(y)\wh W(s)}{(y-s)^2}\ri|\leq \frac {\le|W(y) \wh W(s) \ri|}{\sin^2 \theta|y-b|^2} \label{3-41}\ \ \ \hbox{[see fig. \ref{fig2a}]}.
\ee

which is still jointly integrable w.r.t. the arc lengths (recall that
the directions of approach of $\Gamma$ and $\wh \Gamma$ are such that
the weights tend to zero faster than any power of the local
coordinate). 

\begin{figure}
\epsfxsize 5cm
\epsffile{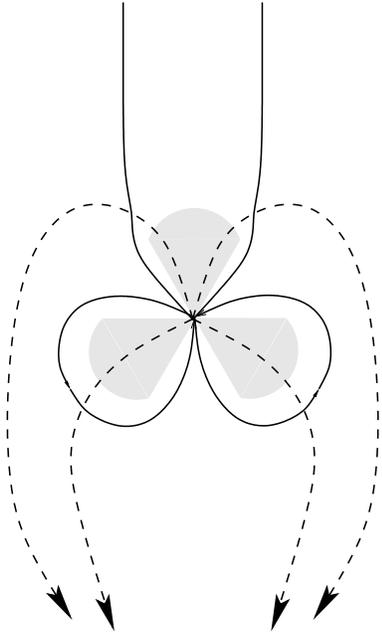}
\caption{\label{dualflowers} 
\fontfamily{cmss}
\fontsize{11pt}{15pt}
\selectfont
The equivalent choice of contours for the dual admissible petals.}
\end{figure}

It is then clear that if $\Gamma$ $\wh \Gamma $ are a circle and a
semi-infinite arc (or vice-versa) the bilinear concomitant for the
corresponding dual solutions is a nonzero constant (which depends on
the choices of the branches of $W$ and $\wh W$). This is immediate for
a pair of contours which intersect only once. For a pair of lassoes
(which intersect twice and with opposite orientations), calling
$p_1,p_2$ the points of intersection we have 
\be
\mathcal B(f_\Gamma,g_{\wh \Gamma}) = \pm (W(p_1)\wh W (p_1)B(p_1) - W(p_2)\wh W(p_2) B(p_2))
\ee
Since the local behavior at the singularity embraced by the lassoes is
a noninteger power, let's say $(y-c)^t$, then the values of $BW\wh W$
on the two intersection points (which lie on different sizes of the
union of the cuts for $W$ and $\wh W$) satisfies 
\be 
W(p_1)\wh W(p_1)B(p_1)  = {\rm e}^{2i\pi t} W(p_2)\wh W (p_2)B(p_2)
\ee
so that 
\be
\mathcal B(f_\Gamma,g_{\wh \Gamma}) = \pm (W(p_1)\wh W (p_1)B(p_1) (1-{\rm e}^{2i\pi t})\neq 0
\label{nonzzz}
\ee
For dual flowers it is convenient to choose different paths for the
dual contours as shown in Fig. \ref{dualflowers}, where the petals
have been replaced by stems using a linear combination of the
contour-integrals of the same petals and stem. The sub-block of the concomitant involving these contours is
nondegenerate, since it can be given a diagonal form with nonzero
entries on the diagonal. The precise values are not important since we
are free to re-scale each solution $f_\Gamma$ and $g_\Gamma$. 
%
Summarizing we have proved that 
\bp
\label{pr31}
There is a normalization of the integrals $f_\Gamma$ and $g_{\wh
  \Gamma}$ such that the bilinear concomitant is precisely the
intersection pairing of the contours $\Gamma$ and $\wh \Gamma$. With
appropriate choice and labeling of the contours the pairing is
represented by the identity matrix. 
\ep
\br
The content of Prop. \ref{pr31} is that if the solutions $f_\Gamma$ and $g_{\wh \Gamma}$ correspond to contours that can be deformed (by Cauchy's theorem and without changing the analyticity properties of the functions $f_\Gamma, g_{\wh \Gamma}$ respectively) in such a way that they do not intersect, then the bilinear concomitant of this pair is zero. 

\begin{figure}
\epsfxsize 9cm
\epsfysize 9cm
\epsffile{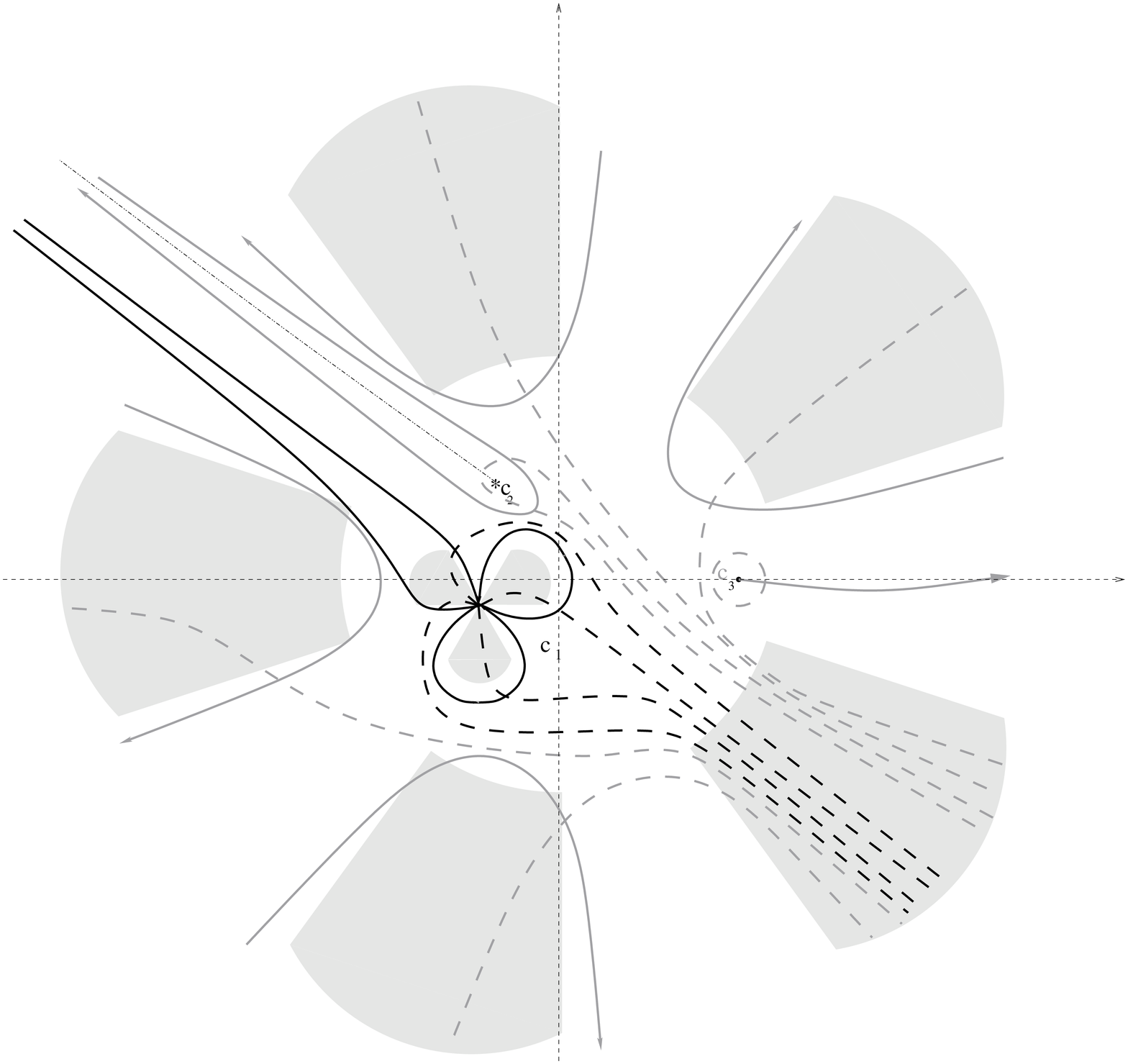}
\caption{The arrangement of dual contours for the same example as in Fig. \ref{figone}: in evidence only the different choice of admissible dual contours at the ``flower''.}
\label{figrem}
\end{figure}

Viceversa, if this cannot be done, the bilinear concomitant is nonzero; we can always choose the contours and dual contours in such a  way that each contour intersects one and only one dual contour. For example the equivalent choice of contours to Fig. \ref{figone} is given by the arrangement in Fig. \ref{figrem}.

Note that two dual lassoes intersect at two points but that --by virtue of (\ref{nonzzz}) their ``weighted'' intersection number is nonzero (whereas the usually defined intersection number would be zero).
\er

%
%
%
\section{Auxiliary wave vectors}
\label{auxwave}
{\bf Caveat} In this section we will make statements concerning the
biorthogonal polynomials $p_n,s_n$ and the corresponding
quasipolynomials $\psi_n,\phi_n$. 
It will  be understood that 
\begin{enumerate}
\item Any statement made on the $\psi_n$'s and the Fourier--Laplace transforms of the $\phi_n$'s admits a specular statement for the $\phi_n$'s and the F-L transforms of the $\psi_n$'s.
\item Any statement made on the $\psi_n$'s admits an analog statement for the $\wh \psi_n$'s and $\wc \psi_n$'s by replacing the moment functional $\L$ with $\wh \L$ or $\wc \L$, and specular statements for $\wh \phi_n,\wc\phi_n$.
\end{enumerate}

Consider the functions
\bea
\mathcal B_2(x;y,s):= \le(\frac {B_2(y)-B_2(s)}{y-s} \le( x - 
\frac 1{y-s}\ri)-\frac{A_2(y) - B_2'(s)-A_2(s)}{y-s}\ri)\\
 \psi_n^{(\wh \Gamma)} := \frac 1{2i\pi} \int_{\wh\Gamma}{\rm d}s 
\Int{\rm d}\xi{\rm d}y\mathcal B_2(x;y,s)
{\rm  e}^{\xi y-xs -  V_2(y)+ \wh V_2(s)} \frac { \psi_n(\xi)}{x-\xi}
  \eea

If $x$ belongs to a contour $\Gamma_{x,\mu}$ of the integration $\Int$ we obtain 
\be
\psi_n^{(\wh \Gamma)}(x)_+ = \psi_n^{(\wh \Gamma)}(x)_- + \sum_{\nu}
\mathcal B_2(\wh \Gamma, \Gamma_{y,\nu}) \varkappa_{\mu,\nu} \psi_n(x)  
\label{jumppsibar}
\ee
where the subscript $x_\pm$ denotes the boundary values from the
left/right and  $\mathcal B_2(\wh \Gamma,\Gamma_{y,\nu})$ stands for the constant (in $x$) bilinear concomitant
\be
\mathcal B_2(\wh \Gamma,\Gamma_{y,\nu}):= \frac 1{2i\pi} \int_{\wh \Gamma} {\rm d}s
\int_{\Gamma_{y,\nu}} {\rm d}y \mathcal B_2(x;y,s) {\rm e}^{\wh V_2(s)-V_2(y)+x(y-s)}\ .
\ee
Therefore
their jump across the contours of discontinuity is a constant multiple
of $ \psi_n(x)$.

We have
\bp
\label{recaux}
The sequences of functions $\{\psi_n^{(\wh\Gamma)}\}_{n\in \N}$ satisfy the same recurrence relations (for $n$ large enough)  as the quasipolynomials $\psi_n$ 
\bea
&& x\le (\psi_n^{(\wh\Gamma)} + \sum_{j=1}^{q_2} \ell_j(n)\psi_{n-j}^{(\wh\Gamma)}\ri)  = \sum_{-1}^{d_2}  \alpha_j(n) \psi_{n-j}^{(\wh\Gamma)} \ ,\qquad n\geq d_2 + q_2\\
&&
\pa_x\le( \psi_n^{(\wh\Gamma)} + \sum_{j=1}^{q_1}  \wc m_j(n+j)\psi_{n+j}^{(\wh\Gamma)}\ri) =   - \sum_{-1}^{d_1}  \wc \b_j(n+j) \psi_{n+j}^{(\wh\Gamma)} \  ,\qquad n\geq 1
\eea
\ep
(For the proof see App. \ref{proofrecaux})
\bd
\label{definewaves}
Beside the wave vector $\Wp$ we define the following $d_2$ auxiliary wave-vectors
\bea
\Wp^{(\nu)}(x):= \frac 1{2i\pi} \int_{\wh\Gamma_{y,\nu}}{\rm d}s 
\Int{\rm d}\xi{\rm d}y\mathcal B_2(x;y,s)
{\rm  e}^{\xi y-xs -  V_2(y)+ \wh V_2(s)} \frac { 1}{x-\xi}\Wp(\xi)\ ,\ \nu=1,\dots,d_2\\
\Wp^{(0)}(x):=\Wp(x)\ .
\eea
We also define the {\bf dual wave vectors}
\bea
&& \unWs^{(0)}(x):= {\rm e}^{V_1(x)}\Int {\rm e}^{\xi y - V_1(\xi)}\frac{1}{x-\xi}\Ws(y) {\rm d}\xi {\rm d}y\\
&&\unWs^{(\mu)}(x):= \int_{\Gamma_{y,\nu}} {\rm d}y \,
{\rm e}^{xy}\Ws(y)\ ,\nu=1,\dots,d_2
\eea
\ed
\bp
\label{recauxdual}
The components of the dual wave vectors satisfy the recurrence relations
\bea
&& x\le(\un \phi^{(\nu)}_n + \sum_{j=1}^{q_2} \wh \ell_j
(n+j)\un \phi^{(\nu)}_{n+j}\ri) = \sum_{j=-1}^{d_2} \wh \a_{j}(n+j)
\un \phi^{(\nu)}_{n+j} + \delta_{\nu 0}\delta_{n0} \sqrt{h_0}{\rm
  e}^{V_1(x)} \ ,\ \ \nu =0,\dots,d_2\\
&& \pa_x\le(\un \phi^{(\nu)}_n + \sum_{j=1}^{q_1} m_{j}(n)
\un \phi_{n-j}^{(\nu)} \ri) = \sum_{j=-1}^{d_1} \b_j(n)\un \phi^{(\nu)}_{n-j} \
,\nu=1,\dots,d_2\ . 
\eea
\ep
\br
The wave vector $\Ws^{(0)}$ does not satisfy a finite-term differential recurrence relation: a formula can be derived but it is not useful for our purposes.
\er
{\bf Proof}
The formul\ae\ for the Fourier--Laplace transforms follow from
integration by parts from the relations satisfied by $\phi_n(y)$
(Prop. \ref{diffrec}). We only point out  that integration by parts
does not give any boundary contribution because $s_n + \sum \wh
\ell_j(n+j)s_{n+j}(y)$ is divisible by $B_2(y)$ and hence vanishes at
the hard-edge end-points. 

The only relation that needs to be checked is the multiplicative relation for $\nu =0$.
Denoting  temporarily by a tilde the linear combination 
\be
\wt \phi_n := \phi_n + \sum_{1}^{q_1} \wh \ell_j(n+j)\phi_{n+j}\ ,
\ee
we have
\bea
x\wt {\un\phi}^{(0)}(x) =&\& {\rm e}^{V_1(x)}\Int {\rm e}^{\xi y - V_1(\xi)}\frac{x}{x-\xi}\wt \phi_n(y) {\rm d}\xi {\rm d}y=\cr
=&\&
{\rm e}^{V_1(x)}\Int {\rm e}^{\xi y - V_1(\xi)} \wt\phi_n(y)  {\rm d}\xi {\rm d}y + {\rm e}^{V_1(x)}\Int {\rm e}^{\xi y - V_1(\xi)}\frac{\xi}{x-\xi}\wt\phi_n(y) {\rm d}\xi {\rm d}y =\cr
=&\&{\rm e}^{V_1(x)}\delta_{n0}\sqrt{h_0}+  {\rm e}^{V_1(x)}\Int {\rm e}^{\xi y - V_1(\xi)}\frac{-\pa_y}{x-\xi}\wt\phi_n(y) {\rm d}\xi {\rm d}y = \cr
=&\& {\rm e}^{V_1(x)}\delta_{n0}\sqrt{h_0}+  \sum_{j=-1}^{d_2} \wh
\a_{j}(n+j)\un \phi^{(0)}_{n+j} \ .\ \ \ \hbox{{\bf Q.E.D.}}
\eea 
\subsection{Christoffel--Darboux identities}
In the general theory of the two--matrix model the following kernel plays an essential r\^ole in the computation of statistical correlation functions
\be
K_{12}^N(x,y):= \sum_{j=0}^{N-1}p_j(x) s_j(y){\rm e}^{-V_1(x)-V_2(y)} = \sum_{j=0}^{N-1} \psi_j(x) \phi_j(y)\ . \label{kernel12}
\ee
In a previous paper by the author and collaborators \cite{BEH, BHI} the
case of polynomial potentials $V_i$ was considered (without
hard-edges) and  it was of capital importance the existence of a
Christoffel--Darboux identity allowing to express $K_{12}^N$ (or
rather some transform of it) in terms of bilinear combinations of the
BOPs involving only a number of BOPs depending only on the degrees of
the potentials.

We look for a similar bilinear expression in this model.

\bd
\label{defwindows}
We define the {\bf windows} of the wave vectors $\Wp^{(\mu)} $ and $\unWs^{(\mu)}$, $\mu = 0,\dots, d_2$
\be
\un \Phi_n^{(\mu)} (x):= [\un \phi^{(\mu)} _{n-1},\dots, \un \phi^{(\mu)}_{n+d_2-1}]\ ,\qquad
\Psi_n^{(\mu)}(x):= [\psi^{(\mu)}_{n-d_2},\dots, \psi^{(\mu)}_{n}]^t\ .
\ee
\ed

We rewrite (\ref{kernel12}) in terms of the wave vectors 
\be
K_{12}^N = \Ws^t(y) \Pi_N\Wp(x)\ ,\qquad \Pi_N:= \le\{\begin{array}{c} 
\delta_{ij}\ ,\ 0\leq i\leq N-1\\
0\ \hbox{ otherwise .}
\end{array}\ri.
\ee
Recall the multiplicative and differential recurrence relations in
Prop. \ref{multrec} and  Prop. \ref{diffrec} (which we rewrite here for
the reader's convenience)  
\bea
\pa_y\Ws^t  (\1 + \wh L)=  - \Ws^t \wh A \ ,\qquad
x (\1 + L) \Wp = A \Wp\cr
(\1 + L)^{-1} A = \wh A(\1 + \wh L)^{-1} =:Q\ .\nonumber
 \eea
Consider now  the following expressions
\bea
(x + \pa_y)&\& \Ws^t(y) (\1 + \wh L) \Pi (\1 + \wh L)^{-1} \Wp(x) = \cr
=&\&
\Ws^t(y) (\1 + \wh L) \Pi (\1 + \wh L)^{-1} \wh A (\1 + \wh L)^{-1} \Wp(x) - \Ws^t(y) \wh A  \Pi (\1 + \wh L)^{-1} \Wp(x) = \cr
=&\&
\Ws^t   \Pi \wh A (\1 + \wh L)^{-1} \Wp   + 
\Ws [\wh L,\Pi] (\1 + \wh L)^{-1}\wh A(\1 + \wh L)^{-1} \Wp +\cr
&&- \Ws^t  \Pi \wh A (\1 + \wh L)^{-1} \Wp  
-\Ws^t  [ \wh A,\Pi] (\1 + \wh L)^{-1} \Wp =\cr
 =&\& \Ws [\wh L,\Pi] \wh Q (\1 + \wh L)^{-1} \Wp - \Ws^t  [ \wh A,\Pi] (\1 + \wh L)^{-1} \Wp =\cr
 = &\& \Ws [\wh L,\Pi] \wh Q \wh \Wp - \Ws^t  [ \wh A,\Pi]  \wh \Wp 
\eea
where we have set $\wh Q:= (\1 + \wh L)^{-1} \wh A$. We now use the
fact that $\wh Q$  is the recurrence matrix for the associated $\wh
\Wp$ wave vector (see Prop. \ref{assorec} where $\wh \Wp := \wh \p
   {\rm e}^{-V_1(x)} $) and obtain 
\be
(x + \pa_y) \Ws^t(y) (\1 + \wh L) \Pi (\1 + \wh L)^{-1} \Wp(x) 
= \Ws(y) [x \wh L-\wh A,\Pi]  \wh \Wp(x) \label{CDI12}
\ee
\bea
&& \wh \Wp = (\1 + \wh L)^{-1} \Wp = \wh \p(x) {\rm e}^{-V_1(x)}\\
&& \wh \Ws = (\1 + \wh L^t) \Ws = \wh \s(y) B_2 {\rm e}^{-V_2(y)} = \wh \s(y) {\rm e}^{-\wh V_2(y)}
\eea
With these notation we have

\bt[Christoffel--Darboux identity]
\label{CDI}
For the kernels 
\bea
\wh  K_{11}^{N,\nu} (x,x') &\& = \int_{\Gamma_{y,\nu}} {\rm e}^{xy}  \wh \Ws^t(y)\Pi_N \wh \Wp(x') = 
{ \wh \unWs^{(j)}(x)}^t \Pi_N\wh  \Wp(x')\ ,\\
K_{11}^{N,\nu} (x,x') &\& = \int_{\Gamma_{y,\nu}}{\rm e}^{xy}   \Ws^t(y) \Pi_N  \Wp(x') = 
{ \unWs^{(j)}(x)}^t \Pi_N  \Wp(x')\ ,\qquad  j=1,\dots ,d_2
\eea
we have the identities
\bea
&& (x'-x)\wh K_{11}^{N,j} (x',x)  =   {\unWs^{(j)}}(x')^t  \m{\Amat}_N(x) \wh \Wp(x)\\
&& (x'-x)K_{11}^{N,j} (x',x)  =  {\unWs^{(j)}}(x')^t  \m{\Amat}_N(x') \wh \Wp(x)\ . 
\eea
(note the argument of $\Amat_N$ in the two formul\ae)
where $\Amat_N(x) :=\le [\wh A - x\wh L,\Pi_N\ri]$.
\et
{\bf Proof}
The identity for $\wh K_{11}^{N,j}(x,x')$ follows by performing 
integration by parts on (\ref{CDI12}) and noticing that the boundary contributions
vanish since $\wh \Phi(y) = B_2(y)\wh \s(y) {\rm e}^{-V_2(y)}$ and $B_2(y)$ vanishes at the hard-edges. 
The identity for $K_{11}^{N,j}(x,x')$ follows from the one for
$\wh K_{11}^{N,j}$ and this manipulation
\bea
(x'-x) \wh \unWs^{(j)}(x')^t\Pi\wh\Wp(x) &\&=(x'-x) \unWs^{(j)}(x')^t(\1 + \wh L)
\Pi\wh\Wp(x) =\cr
&\&=  (x'-x)\bigg(\unWs^{(j)}(x')^t[\wh L,
\Pi] \wh\Wp(x) +\unWs^{(j)}(x')^t\Pi \Wp(x)  \bigg) =\cr
&\& = (x'-x) K_{11}^{N,j} (x',x) + (x'-x) \unWs^{(j)}(x')^t[\wh L,
\Pi] \wh\Wp(x)
\eea
so that 
\bea
&\& (x'-x)  K_{11}^{N,j} (x',x)  = \cr 
&\& =
(x'-x)\wh K_{11}^{N,j} (x',x)
- (x'-x) \unWs^{(j),t} (x')[\wh L,
\Pi] \wh\Wp(x) =  \unWs^{(j),t} (x')  \m{\Amat}_N(x') \wh \Wp(x)
\eea
{\bf Q.E.D.}\par\vskip 5pt
Note that --with a slight abuse of notation-- in the RHS of the CDIs
we can replace the wave vectors $\unWs$ by the corresponding window
$\un\Phi_n$ since the matrix $ \Amat_n$ has a nonzero square block of
size $d_2+1$ with top-right corner in the $(n-1,n)$ entry, and hence
the bilinear expression $\unWs\Amat\Wp$ only involves the terms in the
dual windows $\un \Phi_n$ and $\wh \Psi_n$. We will denote from now on
by $ \Amat$ only the $d_2+1$ square matrix which is relevant to the
pairing.

The importance of the theorem is that we can express the kernel
$K_{11}$ in terms of the dual quantities $\un \phi_n(x)$ and $\wh
\psi_n(x')$ involving only the indexes $N-d_2\leq n \leq N$.  

Note, however, that we must introduce the orthogonal polynomials $\wh
\p$ for the associated moment functional $\wh \L$ in order to find a
Christoffel--Darboux relation similar to the standard one for
orthogonal polynomials.
\bt[Auxiliary CDIs]
\label{auxCDIS}
The auxiliary wave vectors enter in the following auxiliary Christoffel--Darboux identities
\bea
{\bf (a)}\ (z-x) \unWs^{(0)}(z)^t \Pi_n \Wp^{(0)}(x) &\&=\un
\Phi_n^{(0)}(z) \Amat (z) \wh \Psi_n (x) + {\rm e}^{V_1(z)-V_1(x)} \cr 
(z-x) \wh \unWs^{(0)}(z)^t \Pi_n\wh \Wp^{(0)}(x) &\&=\un \Phi_n^{(0)}(z) \Amat (x) \wh \Psi_n (x) + {\rm e}^{V_1(z)-V_1(x)}
\label{aCDI}
\\
{\bf (b)}\ \ (z-x) \unWs^{(j)}(z)^t \Pi_n \Wp^{(k)}(x) &\& = \un
\Phi_n^{(j)}(z) \Amat(z) \wh \Psi_n(x) - \frac 1{2i\pi}
\int_{\Gamma_{y,\nu}} \int_{\wh\Gamma_k} \mathcal B_2(x;y,s) {\rm
  e}^{yz-xs+\wh V_2(s) -V_2(y)} \ ,\cr
  (z-x) \wh \unWs^{(j)}(z)^t \Pi_n \wh \Wp^{(k)}(x) &\& = \un
\Phi_n^{(j)}(z) \Amat(x) \wh \Psi_n(x) - \frac 1{2i\pi}
\int_{\Gamma_{y,\nu}} \int_{\wh\Gamma_k} \mathcal B_2(x;y,s) {\rm
  e}^{yz-xs+\wh V_2(s) -V_2(y)} \ ,\cr
  &\& \ j,k=1,\dots,d_2\ .
  \label{bCDI}
\eea
\et
(For the proof see App. \ref{proofauxCDI}).

\subsection{Ladder matrices}
In this section we derive an expression for the ODE satisfied by the polynomials in terms of the so-called  "folding" (see \cite{BEH}). This will have  certain advantages when explaining the relations between the various ODEs that naturally appear in the problem: a different explicit representation of the ODE will be given in the next section as well, using a completely different approach based upon the explicit integral representations of the wave vectors and on the duality provided by the Christoffel--Darboux pairing.

We first  have the simple lemma
\bl[Ladder matrices]
\label{ladderlemma}
The multiplicative recurrence relations for the wave vectors $\Wp^{(0)}, \unWs = \unWs^{(j)}$ ($j=1,\dots d_2$) 
\bea
&& x(\1 + L)\Wp^{(0)}  = A\Wp^{(0)}\ , \qquad  x  (\1 + \wh L^t)  \unWs^{(j)}= \wh A^t  \unWs^{(j)}
\eea
are equivalent to the relations
\bea
&&\Psi_{n+1}^{(0)} (x) = {\bf a}_n(x)\Psi_n^{(0)}(x)\ ,\label{ladder}\\
&&\un \Phi_{n}^{(j)} (x)  =\un \Phi_{n+1}^{(j)}(x) \un{ {\wh {\A}}}_n(x) \label{ladderdual}
\eea
where 
\def \E{ \le[\begin{array}{c}
0\\ \vdots \\ 0\\1
\end{array}\ri]}
\bea
&& {\bf a}_n(x) = \Lambda -  \frac 1{\a_{-1}(n)} 
\E
[\a_{d_2}(n),\dots, \a_0(n)]
+ \frac {x}{\a_{-1}(n)} \E [0,\dots, \ell_{q_2}(n),\dots, \ell_1(n),1]\\
&& \wh{\un {\bf a}}_n(x) = \Lambda -  \frac 1{\wh\a_{-1}(n\!-\!1)} 
\le[
\begin{array}{c}
\wh\a_{0}(n)\\
\wh\a_{1}{(n\!+\!1)}\\
\vdots\\
\wh\a_{d_2}(n\!+\!d_2)
\end{array}
\ri][1,0,\dots,0]  + \frac {x}{\wh\a_{-1}(n\!-\!1)} \le[
\begin{array}{c}
1\\
\wh\ell_1(n\!+\!1)\\
\vdots\\
\wh\ell_{q_2}(n\!+\!q_2)\\
0\\
\vdots
\end{array}
\ri] [1,0,\dots,0]
\eea
and  $\Lambda$ denotes the upper shift matrix (of size $d_2+1$).
The relations (\ref{ladder}) and (\ref{ladderdual}) hold also for the other sequences of  windows $\Psi_n^{(j)}$ and $\un \Phi_{n}^{(0)}$ provided that $n\geq d_2+q_2$ ($n\geq 1$ respectively).
\el 
{\bf Proof.} The proof follows immediately from  the recurrence relations for the wave vectors $\Wp^{(0)}$ (the quasipolynomials)  and $\unWs^{(j)}$ (the Fourier--Laplace transforms) by solving for  $\psi_{n+1}(x)$ (or $\un\phi_{n-1}$) in terms of $\psi_{n-d_2},\dots,\psi_{n}$ ($\un\phi_{n},\dots,\un \phi_{n+d_2}$) and rewriting the relation in matrix form.
The statement for the other sequences of windows follows from the fact that the corresponding wave vectors satisfy the same finite-term recurrence relations in the specified range (see Prop. \ref{recaux} and Prop. \ref{recauxdual}). 
 {\bf Q.E.D.}\par \vskip 5pt 
\bl[Folded recursion relations]
The differential recurrence relations for the wave-vector $\Wp$ 
\bea
&& \pa_x(\1 + \wc M^t) \Wp = -\wc B^t \Wp
\eea
are equivalent to the relations
\bea
&& \pa_x\bigg(\wc {\mathcal M}_n (x) \Psi_n\bigg) =-\wc {\mathcal B}_n(x) \Psi_n\label{ODEpsi} \\
&&\wc {\mathcal M}_n:= \1 + \sum_{j=1}^{q_1} \wc{\boldsymbol m}_j(n)
 {\bf a}_{n}\cdots {\bf a}_{n+j-1}\label{start1}\\
&&\hspace{1cm} \wc  {\boldsymbol m}_j(n):=   {\rm diag} (\wc m_{j}(n+j-d_2),\dots , \wc m_j(n+j))\\
&&\wc {\mathcal B}_n:= \wc {\boldsymbol  \beta}_{-1}(n) ({\bf a}_{n-1})^{-1}  +\wc {\boldsymbol \beta}_0(n)    +\sum_{j=1}^{d_1} \wc {\boldsymbol \beta}_j(n) {\bf a}_{n}\cdots {\bf a}_{n+j-1}\\
&&\hspace{1cm} \wc  {\boldsymbol \b}_j(n):=   {\rm diag} (\wc \b_{j}(n+j-d_2),\dots , \wc \b_j(n+j))\label{end1}
\eea
\el
{\bf Proof}.
The formula is an iterated application of the ladder recurrence
relations (on a window of consecutive elements with indexes
$n-d_2,\dots, n$) to  the differential recurrence relation for the
wave vector (see \cite{BEH} for more details). {\bf Q.E.D.}\par \vskip 5pt
\br
A completely analogous statement can be derived for the windows of the dual vector $\un\Phi_n^{(j)}$, $j=1,\dots, d_2$.
\er
\br
The matrices ${\bf a}_n$ have a companion-form and are {\bf invertible} since the determinant is $-\frac {\a_{d_2}(n)}{\a_{-1}(n)}$ which has been proved nonvanishing in Thm. \ref{thmrecmul}. Moreover the inverse is also linear in $x$ (the details are left to the reader).
\er
\br
By the very definition $\wc {\mathcal M}_n(x) \Psi_n = \wc \Psi_n$ is
the window of quasipolynomials (and associated functions) for the
moment functional $\wc \L$.
\er
\bc
The $d_2+1$ columns provided by the windows of the auxiliary wave vectors $\Wp^{(j)}(x)$ provide a fundamental system for the ODE (\ref{ODEpsi}) for $n\geq d_2+q_2$.
\ec
{\bf Proof}. From Prop. \ref{recaux} we know that the components of
the auxiliary wave vectors satisfy the same recurrence relations (both
multiplicative and differential) as the quasipolynomials provided $n$
is large enough. Moreover the recurrence relations always involve a
fixed number of terms with indexes "around $n$": since the derivation
of the ODE is entirely based on the recurrence relations the statement
follows. {\bf Q.E.D.}\par \vskip 5pt 
\bp
\label{43}
The determinant of $\wc {\mathcal M}_n(x)$ is proportional to $B_1(x)$ by a nonzero constant.
\ep
{\bf Proof}.
Consider the window of polynomials $\p_n:= [p_{n-d_2},\dots,p_n]^t$: from the definition of the matrix $\wc M$ it follows that 
\be
\wc {\mathcal M}_n(x) \p_n(x) = B_1(x) \wc \p_n(x)  
\ee
We first prove that $\det \wc{\mathcal M}_n$ (which is {\em a fortiori} a polynomial) is divisible by $B_1$.
Let $c$ be a zero of $B_1$ of multiplicity $r$: at least one component (say the $\ell$-th) of $\p_n(c)$ is nonzero because of the very  genericity assumption which guarantees the existence of $\wc M$ (\ref{gencheck}). Let $E(x)$ be the matrix obtained by replacing the $\ell$-th column of the identity with $\p_n(x)$. Clearly $\det E(x)$ is nonzero in a neighborhood of $x=c$ by our definition of $\ell$.
It follows that the $\ell$-th column of $ \wc {\mathcal M}_n E$ is precisely $B_1\wc \p_n$ and hence each component vanishes at $c$ of order $r$. Also 
\be
\det \wc {\mathcal M}_n E = p_{n-d_2+\ell-1}(x)\det \wc {\mathcal M}_n
\ee
and $p_{n-d_2-1+\ell}(c)\neq 0$. On the other hand $\det \wc {\mathcal M}_n E$ must vanish at $x=c$ of order $r$ since the whole $\ell$-th column does. Repeating this for all roots of $B_1$ we find the assertion of divisibility of $\det \wc{\mathcal M}_n$ by $B_1(x)$. 

On the other hand, using a technique of evaluation of determinants used in \cite{BEH}, 
\bea
\det \wc{\mathcal M}_n = \det \le[\1_{(d_2+1)(q_1+1)} - \le[
\begin{array}{c|c|c|c}
 & {\bf a}_{n+q_1} & & \\
 \hline 
 &&\ddots&\\
 \hline
 &&&{\bf a}_{n}\\
 \hline
 \wc  {\boldsymbol m}_{q_1}(n)&\cdots& \wc  {\boldsymbol m}_{1}(n)&0
\end{array}
\ri]\ri] 
\eea
Considering carefully the structure of the sparse matrix in the last identity, one realizes that the highest power in $x$ is 
\be
\det \wc {\mathcal M}_n = x^{q_1} \frac {\wc m_{q_1}(n+q_1)}{\prod_{j=1}^{q_1} \a_{-1}(n+j)}  + \mathcal O(x^{q_1-1})
\ee
This shows that (since the coefficient does not vanish as per
(\ref{nonzero},\ref{nonzerodual})) the determinant is of degree $q_1 =
\deg B_1$; since it must be also divisible by $B_1$, this concludes
the proof. {\bf Q.E.D.}\par \vskip 5pt
\bc
The windows $\Psi_n, \ \wc \Psi_n$ satisfy
\bea
&& \pa_x \Psi_n = - {\wc {\mathcal M}_n}^{-1}\bigg(  \wc {\mathcal B}_n   + \pa_x {\wc {\mathcal M}_n}\bigg)\Psi_n\\
&& \pa_x \wc \Psi_n =- \wc {\mathcal B}_n {\wc {\mathcal M}_n}^{-1} \wc \Psi_n\label{checkde}
\eea
where $\wc{\mathcal B}_n, \wc{\mathcal M}_n$ are defined in ((\ref{start1})--(\ref{end1})). The ODEs have the same singularity structure as $V_1'$.
\ec
The first relation follows from (\ref{ODEpsi}) and the second from the fact that $\wc {\mathcal M}_n(x) \Psi_n(x) =\wh \Psi_n(x)$.

This shows that the ODE's for $\Psi_n$ and $\wc \Psi_n$ are gauge-equivalent, the gauge being provided by the (polynomial) matrix $\wc {\mathcal M}_n$.
Moreover formula (\ref{checkde}) together with Prop. \ref{43}  shows that the singularities of the
differential equation are at the zeroes of $B_1(x)$.

\subsection{Differential equations for the dual pair of systems}
In this section we present an explicit formula for the ODE satisfied
by the dual pair of fundamental systems, in particular the polynomials
$\wh \psi_n$ and the Fourier--Laplace transforms $\un\phi_n$'s.
 The result generalizes those of \cite{BE}  but the method of derivation is similar to the one adopted in \cite{BHI},
with additional complications deriving from the presence of 
boundary contributions in the integration by parts at various steps of
the derivation.

\paragraph {Notation.} In the proof of this and the following theorems we will encounter
integrations by parts that yield nonzero boundary
contributions. Typically we will encounter  integrals of the form
\be
 \Int y{\rm e}^{\rho y-V_1(\rho)} F(\rho) \phi_m(y){\rm d}y {\rm d}\rho\ ,
\ee 
where $F(\rho)$ is some expression (typically polynomial or rational
in $\rho$)  possibly depending on ``external''
variables. If we attempt an integration by parts on the term $y{\rm
  e}^{y\rho} = \pa_\rho{\rm e}^{y\rho}$, we obtain a certain number of
boundary terms. In all cases they will be boundary evaluation on the
various contours $\Gamma_{x,\mu}$; it is the nature of all these
integrals that only the contours emanating from a {\bf hard--edge}
point give a contribution,  due to the fast decay of ${\rm
  e}^{-V_1(\rho)}$ at all the boundary points of the other contours.
In the above example and in all minute detail, we have
\bea
 \Int y{\rm e}^{\rho y-V_1(\rho)} F(\rho) \phi_m(y) = -\Int {\rm
   e}^{\rho y-V_1(\rho)} (-\pa_\rho + V_1'(\rho)) F(\rho) \phi_m(y) +
 \hbox{(Boundary terms)}\cr
\hbox{(Boundary terms)} = 
\sum_{\mu=1}^{d_1}{\rm e}^{-V_1(\rho)} F(\rho)
 \sum_{\nu=1}^{d_2} \varkappa_{\mu,\nu}\int_{\Gamma_{y,\nu}} \hspace{-15pt}{\rm e}^{\rho y }\phi_m(y) \bigg|_{\rho\in \pa \Gamma_{x,\mu}}
\eea
The evaluation at the boundary points of  the various contours $\Gamma_{x,\mu}$ is clearly to be
understood as limits along the contours; the decay of ${\rm
  e}^{-V_1(\rho)}$ along the contours gives zero contributions except
for the hard--edge contours, at the (finite) boundary of which
$V_1(\rho)$ is regular. In order to economize on space, we introduce
the following shorthand notation for the above boundary terms
\bea
F(\rho){\rm e}^{-V_1(\rho)} \un \phi^{(\varkappa)}(\rho)\bigg|_{\rho
  \in \pa_x \varkappa}:= 
\hbox{(Boundary terms)} 
\eea

\bt
\label{ODEdual}
The dual fundamental system.
\be
\un{ \boldsymbol \Phi}_n(x):=  \le[{\Phi_n^{(0)}\atop {\ds \vdots \atop \ds \Phi_{n}^{(d_2)}}}  \ri]=\le[
\begin{array}{cccc}
\phi_{n-1}^{(0)} & \phi_{n}^{(0)}  & \dots& \phi_{n+d_2-1}^{(d_2)}\\
\hline
\phi_{n-1}^{(1)} & \phi_{n}^{(1)}  & \dots& \phi_{n+d_2-1}^{(d_2)}\\
\vdots & &  & \vdots\\
\phi_{n-1}^{(d_2)} & \phi_{n}^{(d_2)}  & \dots& \phi_{n+d_2-1}^{(d_2)}
\end{array}
\ri] 
\ee
satisfies the ODE
\bea
\un{ \boldsymbol\Phi}_n^{-1} (x)\un{ \boldsymbol\Phi}_n'(x) &\&= \le[\begin{array}{ccccc}
V_1' (x)& 0  & \dots  && 0 \\
P_{n,n-1}  & P_{n,n} & \dots && P_{n,n+d_2-1}\\
0& P_{n+1,n} & & & \vdots\\
0&0&\ddots &&\\
0&0&0&P_{n+d_2,n+d_2-1} &P_{n+d_2-1,n+d_2-1}
\end{array}
\ri]  +\cr
&\&+  {\rm diag} (P_{n+d_2,n-1},\dots ,P_{n+d_2,n+d_2-1}) {\un \A_{n} }^{-1}(x)  + \Amat(x)\le[\frac {\wh \Psi_n(\rho) \un \Phi_n^{(\varkappa)} (\rho) } {x-\rho} \ri]_{\rho\in \pa_x \varkappa} - \Amat(x) W(x)\cr 
&\& W_{ab} (x):= \L\le(\wh p_{n-d_2+a}(\rho) \frac {V_1(\rho) - V_1(x)}{\rho - x}\bigg| s_{n-1+b}(y)\ri)\ ,\ a,b=0,1,\dots,d_2\\
&\& P_{j,k}:= \L (p_j | y s_k)\ .
\eea
where $\un \A_{n}$ is the ladder matrix for the dual wave vector 
(Note that $P = ((1 + M)^{-1} B)^t$)
\et
(For the proof see App. \ref{proofODEdual}).
\bt
\label{ODEdirhat}
The direct system
\bea
\boldsymbol{ {\wh \Psi}}_n(x) :=  \le[\wh \Psi_{n}^{(0)}| \wh\Psi_{n-1}^{(1)}  \cdots  \wh\Psi_{n}^{(d_2)} \ri]=\le[
\begin{array}{c|ccc}
\wh\psi_{n-d_2}^{(0)} & \wh\psi_{n-d_2}^{(1)}  & \dots&\wh \psi_{n-d_2}^{(d_2)}\\
\vdots & &  & \vdots\\
\wh\psi_{n-1}^{(0)} &\wh \psi_{n-1}^{(1)}  & \dots&\wh \psi_{n-1}^{(d_2)}\\
\wh\psi_{n}^{(0)} & \wh\psi_{n}^{(1)}  & \dots& \wh\psi_{n}^{(d_2)}
\end{array}
\ri] 
\eea
 satisfies the  ODE
 \bea
 \boldsymbol{ {\wh \Psi}}_n'\, \boldsymbol{ {\wh \Psi}}_n^{-1} = &\&
-  \le[
 \begin{array}{ccccc}
 \wh P_{n-d_2, n-d_2} &&\dots &\wh P_{n-d_2,n-1} &0\\
 \wh P_{n-d_2+1,n-d_2}&&&\vdots & 0 \\
 0&   \ddots &&&\vdots \\
  0& &\wh P_{n-1,n-2} &\wh P_{n-1,n-1} &0\\
 &&&\wh P_{n,n-1} &V_1'(x)
 \end{array}
 \ri] + {\rm diag} (\wh P_{n+1,n-d_2},\dots, \wh P_{n+1,n}) \wh {\A}_{n-1}^{-1}  +\cr
 &\&
 +\le[ \frac {\wh \Psi_n(\xi) \un \Phi_n^{(\varkappa)}(\xi) } {\xi -x} \ri]_{\xi \in \pa_x \varkappa}  \Amat(x)  + W(x) \Amat(x) \cr
 &\& \wh P_{j,k} := \wh \L(\wh p_j | y\wh  s_k)\ ,
  \eea
  where $W(x)$ was defined in the previous theorem and $\wh {\A}_{n-1} $ is the ladder matrix implementing the multiplicative recurrence relations $\wh\Psi_{n} = \wh {\A}_{n-1}  \Psi_{n-1} $ as per Lemma \ref{ladderlemma} (in particular eq. (\ref{ladder})) specified to the hat-wave vectors. 
\et
(For the proof see App. \ref{proofODEdirhat}).
\section{Dual Riemann--Hilbert problems}
\label{dualRH}
The shape of the Christoffel--Darboux identity (Thm. \ref{CDI}) suggests that the duality of the Riemann--Hilbert problems (and of the differential equations) involves naturally the dual pair of fundamental systems  $\ds \un{\boldsymbol \Phi}_n(x), \wh{\boldsymbol\Psi}_n(x)$ defined in Thm. \ref{ODEdual} and Thm. \ref{ODEdirhat}. 
Recall (from Section \ref{bilconco})  that we can choose a basis in the relative homology of contours $\Gamma_{y,\nu}$ and $\wh \Gamma_{y,\nu}$ (and a rescaling of the $\wh \Wp^{(j)}$ wave vectors depending only on the residues of $V_2'(y){\rm d}y$)  which span the solution space of the two adjoint equations and with bilinear concomitant  
\be
\mathcal B_2(\Gamma_{y,\nu},\wh \Gamma_{y,\mu}) := \Gamma_{y,\nu}\sharp \wh \Gamma_{y,\mu} = \delta_{\mu\nu}\ .
\ee
We can rewrite (Thm. \ref{CDI}) as 
\bea
(x-x')\sum_{j=0}^{n-1} \un {\wh \phi}_j^{(\nu)} (x) \wh \psi_j^{(0)}  (x') =\un{   \Phi}_n^{(\nu)}(x) \Amat (x')\wh { \Psi}_n^{(0)} (x')\label{hatCDI} \\
(x-x')\sum_{j=0}^{n-1} \un {\phi}_j^{(\nu)} (x) \psi_j^{(0)} (x') =\un{   \Phi}_n^{(\nu)} (x) \Amat (x)\wh { \Psi}^{(0)}_n(x')\label{regCDI} 
\eea
$\nu=1,\dots, d_2$, 
where we stress the fact that on the LHS we have the quasipolynomials $\psi_n$ whereas on the RHS we have the $\wh \psi_n$'s.

\bt
\label{perfect}
The fundamental dual pair is put in {\bf perfect duality} by the Christoffel--Darboux matrix $\Amat$
\be
\un{ \mathbf \Phi}_n(x) \Amat_n(x) \wh {\mathbf  \Psi}_n(x) =
\le[\begin{array}{c|c}
1&0\\
\hline
0& \mathcal B_2(\bullet,\bullet)
\end{array}\ri]
\ee
where $\mathcal B_2(\bullet,\bullet)$ represents the (constant in $x$)
bilinear
concomitant for the solutions of the adjoint ODEs along the contours
$\Gamma_{y,\nu}$, $\wh\Gamma_{y,\mu}$, $\mu,\nu = 1,\dots, d_2$. By suitable choice of the
homology classes we have seen that we can always assume it to be
diagonal. The entries on the diagonal  are nonzero and may be set to
$1$ by suitable rescaling of the $d_2$ left-most columns of
$\boldsymbol \Psi_n$: these re-scalings depend on the way we have
performed the cuts in the definitions of $V_2$ and $\wh V_2$ but
depend {\bf only} on the residues of $V_2'$ mod $\Z$. 
\et
(For the proof see App. \ref{proofperfect}).

\subsection{Riemann--Hilbert data}
In this section we summarily indicate how to obtain the data of the
Riemann--Hilbert problems solved by the dual fundamental systems. The
details are considerably involved and not strictly necessary in this
paper. They will appear in a different publication.

Since the two matrices $\un {\boldsymbol \Phi}_n$ and $\wh {\boldsymbol \Psi}_n$
are put in perfect duality by the Christoffel--Darboux pairing, it is
-in principle- sufficient to describe the Riemann--Hilbert data of one
of the two members of the pair, the data for its partner being
completely determined by duality.

It is significantly simpler to analyze the RH data for the matrix
$\un{\boldsymbol \Phi}_n$. We recall that this means controlling the
jump discontinuities and the asymptotic behaviors near the
singularities.

{\bf Jump discontinuities.} They are uniquely due to the first row in the
definition of $\un{\boldsymbol \Phi}_n$ and occur at the contours
$\Gamma_{x,\nu}$:
\be
\un{\boldsymbol \Phi}_n(x_+) =
\le[\begin{array}{ccccc}
1& 2i\pi \varkappa_{\nu,1} &2i\pi \varkappa_{\nu,2} & \dots & 2i\pi
\varkappa_{\nu,d_2}\\
 & 1 & \\
&&\\
&&\ddots\\
\\
\\
&&&&1
\end{array}\ri] \un{\boldsymbol \Phi}_n(x_-)
\ee
where $x_\pm$ denote the boundary values on the left/right of the
point $x\in \Gamma_{x,\nu}$.

Note that the fundamental matrix $\wh {\boldsymbol \Psi}_n(x)$ satisfies a similar jump condition which can be read off eq. (\ref{jumppsibar}) (specified to the $\wh \psi_n$ quasipolynomials).

{\bf Singularities}
The bottom $d_2$ rows (the Fourier--Laplace transforms) are entire
functions. The only singularities in the finite part of the plane
arise from the first row $\un \Phi_n^{(0)}(x)$: apart from the jump
discontinuities (discussed above) we have all the singularities of
${\rm e}^{V_1(x)}$ and the logarithmic branching singularities around the
hard-edge endpoints. Note that the (piecewise analytic) function
\be
F_n(x):= \Int \vec \Phi_n(y)\frac{{\rm e}^{-V_1(\xi) + \xi y}}{x-\xi} =
{\rm e}^{-V_1(x)} \un \Phi_n(x)
\ee
has a well defined limit as $x$ approaches any of the non hard-edge
endpoints (where it is understood that the approach occurs 
within one connected component of its domain of analyticity). Indeed,
if $c$ is such a point one finds 
\be
F_n(c) = \Int \vec \Phi_n(y)\frac{{\rm e}^{-V_1(\xi) + \xi y}}{c-\xi} 
\ee 
which is a well-defined value.
In other words, near a non hard-edge singularity one has 
\be
\un{\boldsymbol \Phi}_n(x) \sim {\rm diag}\le({\rm e}^{V_{1,sing}(x)},
1,\dots,1\ri)Y_0(\1 + \mathcal O(x-c)).
\ee
where $Y_0$ is just the evaluation of the Fourier--Laplace rows and
the $F_n(x)$ defined above at the point $c$, and $V_{1,sing}$ denotes
the singular part of $V_1$ at $c$.

Near a hard--edge point $x=a$, if $\Gamma_{x,\nu_a}$ is the 
the hard-edge contour originating from $a$, we find that the matrix
\be
Y(x) := \le[\begin{array}{cccc}
1 & \ln(x-a) \varkappa_{\nu_a,1} & \dots & \ln(x-a)
\varkappa_{\nu_a,d_2}\\
&&\\
&\ddots\\
&&&\\
&&&1
\end{array}
\ri]\un {\boldsymbol \Phi}_n(x)
\ee
has a removable singularity at $x=a$ and from this we can obtain the
asymptotic behavior near the hard--edge endpoints. 

{\bf Stokes Phenomenon.} 
Possibly the most intricate part is the description of the Stokes'
phenomenon at $x=\infty$.

Indeed, apart from the aforementioned jump-discontinuities of $\un
\Phi_n^{(0)}$  in a neighborhood of $\infty$ (which may be interpreted
as part of the Stokes data), the first row displays no Stokes'
phenomenon, and has an asymptotic behavior which encodes the
orthogonality
\be
\un\phi_n^{(0)}(x) = {\rm e}^{V_1(x)} \Int {\rm
    e}^{-V_1(\xi)+\xi y}\frac {\phi_{n}(y)}{x-\xi} \sim \sqrt{h_n}
  {\rm e}^{V_1(x)} x ^{-n-1} (1 + \mathcal O(1/x))
\ee
The remaining part of the Stokes phenomenon is given by the asymptotic
behavior of the $d_2$ Fourier--Laplace transforms: this is precisely
the same Stokes' phenomenon displayed by the solutions of the ODE
\be
\big(A_2(\pa_x) -x B_2(\pa_x)\big)f =0
\ee
These solutions are described by contour integrals of the same kind as
the ones appearing in the expressions for $\un \Phi_n^{(\nu)}$; a
standard steepest descent formal argument shows that the leading
asymptotic is determined by the saddle-point equation
\be
\frac {A_2(y) +B_2'(y)}{B_2(y)}=V_2'(y) =x\label{saddle}
\ee
($x\to \infty$)
which has $d_2-H$ solutions ($H$ being the number of hard-edge
contours, i.e. the number of (simple) zeroes of $B_2$ which cancel
against corresponding zeroes of the numerator in (\ref{saddle})).

Whereas it is not very difficult to analyze the formal properties of
the asymptotic, it is considerably harder and outside of the intents
of the present paper to present the Stokes matrices associated to this
Stokes' phenomenon. We leave this topic to a different publication.
\subsubsection{Isomonodromic deformations}
The (generalized) $2$-Toda equations for this reduction as explained
in the introduction, determine the evolution of the biorthogonal
polynomials under infinitesimal deformations of the parameters
entering the semiclassical data $A_i,B_i$. It is more convenient to
parametrize the polynomials $A_i, B_i$ not by their coefficients but
by the location of the zeroes of $B_i$ and the coefficients in the
partial fraction expansions of the derivative potentials
$V'_i$. Following the strategy in our \cite{BEH,  BEHiso, BeGe} one
could easily write the pertinent $2$-Toda flows corresponding to these
infinitesimal deformations. 

At the level of the pair of fundamental systems the flows will
generate isomonodromic deformations for the  ODEs satisfied by
$\un{\boldsymbol \Phi}_n$ and $\wh {\boldsymbol \Psi}_n$, provided
that the exponents of formal monodromy at the singularities remain
unchanged. In this case these are precisely the residues of
$V_1'(x){\rm d}x$ and $V_2'(y){\rm d}y$ at the various singularities. 

The reason why the deformations are isomonodromic is that --by their very definition--
the fundamental systems are functions of these deformation parameters
and  the matrices $\dot {\un {\boldsymbol \Phi}}_n
{\un{\boldsymbol\Phi}_n}^{-1}$ (and $\dot {\wh {\un {\boldsymbol
      \Psi}}}_n {\wh {\un{\boldsymbol\Psi}}_n}^{-1}$, the dot
representing a derivative w.r.t. one of the monodromy-preserving
parameters) are rational (or polynomial)  functions of $x$,  which
follows from the analysis of their behavior at the  various
singularities (\cite{JMU, BeMo} for details on the general properties
of isomonodromic deformations). 

The details of this isomonodromic system could be derived from the
complete Riemann--Hilbert characterization of the fundamental systems
and are beyond the scope of this paper, although their derivation is
-in principle- a straightforward computation.

\appendix
\section{Proofs}
In this appendix we report all proof of more technical nature. The expressions are rather long and hence to shorten them we have decided to suppress explicit reference to the variables of integration in the multiple integrals below, since which variables are integrated on which contour is unambiguously implied by the context. We have adhered to the following  general naming scheme: the variables $\xi, \rho$ are integrated along the contours $\Gamma_{x,\nu}$ appearing in the integral $\Int$, the variables $y$ and $\eta$ are variables integrated on the $\Gamma_{y,\mu}$'s. The variable $s$ is always running along the dual contours $\wh \Gamma_{y,\mu}$ (the admissible contours for the differential $\wh W(s){\rm d}s = {\rm e}^{\wh V_2(s)}{\rm d}s = \frac{{\rm e}^{V_2(s)}}{B_2(s)}{\rm d}s$).
\subsection{Proof of Proposition \ref{recaux}}
\label{proofrecaux}
We temporarily denote by a tilde the following linear combination
\be
\wt \psi_n = \psi_n + \sum_1^{q_2} \ell_j(n) \psi_{n-j}
\ee
and notice that 
\be
x\wt \psi_n = \sum_{-1}^{d_2} \a_j(n)\psi_{n-j}\ .
\ee
For the transformed functions $\psi_n^{(\wh \Gamma)}$ (denoting by a tilde the same linear combination) 
\bea
&&  x \wt \psi_n^{(\wh \Gamma)}  =
\frac x{2i\pi } \int_{\wh\Gamma}\int\!\!\!\int_{\varkappa} \mathcal B_2(x;y,s) {\rm
  e}^{\xi y-xs -  V_2(y)+ \wh V_2(s)} \frac {\wt \psi_n(\xi)}{x-\xi}
  = \\
&& =
\frac 1{2i\pi} \int_{\wh\Gamma}\int\!\!\!\int_{\varkappa} \mathcal B_2(x;y,s) {\rm
    e}^{\xi y-xs - V_2(y)+ \wh V_2(s)}\le(  \wt \psi_n(\xi) +\frac
  {\xi \wt \psi_n(\xi)}{x-\xi}\ri)  
  = \\
&&=\sum_{-1}^{d_2}  \alpha_j(n)  \psi_{n-j}^{(\wh\Gamma)}  +
  \underbrace{\int_{\wh\Gamma}\int\!\!\!\int_{\varkappa} \mathcal
    B_2(x;y,s) {\rm 
    e}^{\xi y-xs -  V_2(y)+ \wh V_2(s)} \wt \psi_n(\xi)}_{=0 \hbox {
      for } n\geq  d_2+ q_2}
\eea
where the last term vanishes for $n\geq q_2+d_2$ because the bilinear concomitant kernel $\mathcal B_2(x;y,s)$ is a polynomial in $y$ of degree $d_2-1$ and the linear combination $\wt \psi_n$ contains the orthogonal function $\psi_{n-q_2}$. 

For the differential equation we have (by definition of the $\wc \psi_n$'s)
\be
\wc \psi_n := \psi_n + \sum_{1}^{q_1} \wc m_j(n+j) \psi_{n+j}
\ee
We  then have
\bea
 \pa_x {\wc \psi_n^{(\wh\Gamma)}} =&\&
\int_{\wh\Gamma}\int\!\!\!\int_{\varkappa}
{\rm e}^{-xs} (\pa_x-s)\frac{
 \mathcal B_2(x;y,s) }{x-\xi} {\rm
  e}^{\xi y -  V_2(y)+ \wh V_2(s)}  {{\wc
  \psi_n}(\xi)} 
  = \cr
=&\&
\int_{\wh\Gamma}\int\!\!\!\int_{\varkappa}
{\rm e}^{-xs} \frac{B_2(y)-B_2(s) }{y-s} {\rm
  e}^{\xi y -  V_2(y)+ \wh V_2(s)}  \frac {{\wc
  \psi_n}(\xi)} {x-\xi} + \cr
  &\&  + 
 \int_{\wh\Gamma}\int\!\!\!\int_{\varkappa} {\rm
  e}^{\xi y-xs -  V_2(y)+ \wh V_2(s)}  {{\wc
  \psi_n}(\xi)} 
 (-\pa_\xi-s) \frac{
 \mathcal B_2(x;y,s) }{x-\xi} 
  =\cr
  \m{=}^{\star}&\&
  \int_{\wh\Gamma}\int\!\!\!\int_{\varkappa}
\frac{B_2(y)-B_2(s) }{y-s} {\rm
  e}^{\xi y -xs -  V_2(y)+ \wh V_2(s)}  \frac {{\wc
  \psi_n}(\xi)} {x-\xi} + \cr
  &\&  + 
 \int_{\wh\Gamma}\int\!\!\!\int_{\varkappa}\frac{
 \mathcal B_2(x;y,s) }{x-\xi}  (\pa_\xi-s) {\rm
  e}^{\xi y-xs -  V_2(y)+ \wh V_2(s)}  {{\wc
  \psi_n}(\xi)} 
  =\cr
=&\& -\sum_{-1}^{d_1} \wc \b_j(n+j) \psi_{n+j}^{(\wc \Gamma)} +\cr
&\&+
\int_{\wh\Gamma}\int\!\!\!\int_{\varkappa}
\frac{B_2(y)-B_2(s) }{y-s} {\rm
  e}^{\xi y -xs -  V_2(y)+ \wh V_2(s)}  \frac {{\wc
  \psi_n}(\xi)} {x-\xi} + \cr
  &\&  + 
 \int_{\wh\Gamma}\int\!\!\!\int_{\varkappa}  \mathcal B_2(x;y,s) (y-s) {\rm
  e}^{\xi y-xs -  V_2(y)+ \wh V_2(s)} \frac  {{\wc
  \psi_n}(\xi)} {x-\xi} = \cr
  =&\&
  -\sum_{-1}^{d_1} \wc \b_j(n+j) \psi_{n+j}^{(\wc \Gamma)} +\cr
&\&+
\int_{\wh\Gamma}\int\!\!\!\int_{\varkappa}
\big(x(B_2(y)-B_2(s)) - A_2(y) + B_2'(s)+A_2(s) \big) {\rm
  e}^{\xi y -xs -  V_2(y)+ \wh V_2(s)}  \frac {{\wc
  \psi_n}(\xi)} {x-\xi} \m{=}^{\hbox{ \tiny (the $s$-part is a total derivative)}}  \cr
 = &\&  -\sum_{-1}^{d_1} \wc \b_j(n+j) \psi_{n+j}^{(\wc \Gamma)} +
\int_{\wh\Gamma}\int\!\!\!\int_{\varkappa}
\big(xB_2(y) - A_2(y) \big) {\rm
  e}^{\xi y -xs -  V_2(y)+ \wh V_2(s)}  \frac {{\wc
  \psi_n}(\xi)} {x-\xi} = \cr
=&\&   -\sum_{-1}^{d_1} \wc \b_j(n+j) \psi_{n+j}^{(\wc \Gamma)} +
\int_{\wh\Gamma}\int\!\!\!\int_{\varkappa} B_2(y) {\rm
  e}^{\xi y -xs -  V_2(y)+ \wh V_2(s)}  {\wc
  \psi_n}(\xi) +\cr
 &\&+\overgroup{\int_{\wh\Gamma}\int\!\!\!\int_{\varkappa}
\big(\xi B_2(y) - A_2(y) \big) {\rm
  e}^{\xi y -xs -  V_2(y)+ \wh V_2(s)}  \frac {{\wc
  \psi_n}(\xi)} {x-\xi}}^{\hbox{\tiny total derivative in $y$}} = \cr
=&\&   -\sum_{-1}^{d_1} \wc \b_j(n+j) \psi_{n+j}^{(\wc \Gamma)} +
\undergroup{\int_{\wh\Gamma}\int\!\!\!\int_{\varkappa} B_2(y) {\rm
  e}^{\xi y -xs -  V_2(y)+ \wh V_2(s)}  {\wc
  \psi_n}(\xi)}_{=0 \hbox{ for } n\geq q_2+1} 
\eea
In the step marked with $\star$ we have  performed an integration by parts: in this integration we do not get any boundary  contributions because the quasipolynomials $\wc\psi_n$ by definition are divisible by $B_1$ (which vanishes at all endpoints and in particular at the hard-edge ones).
This concludes the proof. {\bf Q.E.D.}\par \vskip 5pt 
\subsection{Proof of Theorem \ref{auxCDIS}}
\label{proofauxCDI}
During this and following  proofs we use the notation 
\be
\vec \Phi_n(y):= [\phi_{n-1},\dots, \phi_{n+d_2-1}]\ ,\label{notat}
\ee
for the row-vector of quasipolynomials in $y$.
Moreover, at the risk of marginal confusion, we omit all differentials
of the integration variables since which variables are integrated and
on which contour should be  always uniquely determined by the context
(the formulas become significantly longer otherwise).
For eq. (\ref{CDI})  we have (recall that $\Amat(\xi)$ is linear in $\xi$)
\bea
\hbox {(LHS of \ref{aCDI})} =&\& \sum_{j=0}^{n-1} {\rm e}^{V_1(z)} \Int {\rm
  e}^{-V_1(\xi)+\xi y} \frac {\phi_j(y)}{z-\xi} \psi_j(x) (z-x) =\cr 
=&\& {\rm e}^{V_1(z)} \Int {\rm e}^{-V_1(\xi) + \xi y}\vec  \Phi_n (y)
\Amat (\xi) \wh \Psi_n(x) \frac{z-x}{(z-\xi)(\xi-x)} = \cr 
=&\& {\rm e}^{V_1(z)} \Int {\rm e}^{-V_1(\xi) + \xi y} \vec \Phi_n (y)
\Amat (\xi) \wh \Psi_n(x) \le(\frac 1{z-\xi} - \frac 1{x-\xi} \ri)
=\cr 
=&\& {\rm e}^{V_1(z)} \Int {\rm e}^{-V_1(\xi) + \xi y}\frac{ \vec \Phi_n (y)}{z-\xi}  \Amat (\xi) \wh \Psi_n(x) -  
{\rm e}^{V_1(z)} \Int {\rm e}^{-V_1(\xi) + \xi y} \vec \Phi_n (y)  \Amat (\xi)\frac { \wh \Psi_n(x)}{x-\xi} =\cr
\m{=}^{\star}&\& {\rm e}^{V_1(z)} \Int {\rm e}^{-V_1(\xi) + \xi
  y}\frac{ \vec \Phi_n (y)}{z-\xi}  \Amat (z) \wh \Psi_n(x) -   
{\rm e}^{V_1(z)} \Int {\rm e}^{-V_1(\xi) + \xi y} \vec \Phi_n (y)  \Amat (x)\frac { \wh \Psi_n(x)}{x-\xi} =\cr 
=&\& \un \Phi_n^{(0)}(z) \Amat(z) \wh \Psi_n(x)  + {\rm
  e}^{V_1(z)}\sum_{j=0}^{n-1}  \Int {\rm e}^{-V_1(\xi) + \xi y}
\wh\phi_j (y)  \wh \psi_j(x) =\cr 
=&\&\un \Phi_n^{(0)}(z) \Amat(z) \wh \Psi_n (x) + {\rm e}^{V_1(z)-V_1(x)}\ ,
\eea
where in the identity marked $\star$ we have used the linearity of $\Amat$ which implies the following identity
\be
\frac{\Amat(\xi)}{z-\xi}  - \frac {\Amat(\xi)}{x-\xi} = \frac {\Amat(z)}{z-\xi} - \frac {\Amat(x)}{x-\xi}\ .
\ee
The second form of ${\bf (a)}$ is proved along the same lines using
the principal CDI for the kernel $\wh K_{11}$ (in Thm. \ref{CDI}).
For the remaining CDI's  we have 
\bea
\hbox {(LHS of \ref{bCDI})}  =&\& \frac {z-x}{2i\pi}\sum_{r=0}^{n-1} \int_{\Gamma_j}{\rm
  e}^{zy} \phi_r(y) \int_{\wh \Gamma_k}\Int \mathcal B_2
  (x;\eta,s) {\rm e}^{\eta \rho -xs + \wh
  V_2(s)-V_2(\eta) }
   \frac {\psi_j(\rho)}{x-\rho} = \cr 
=&\& \frac 1{2i\pi} \int_{\Gamma_j}{\rm e}^{zy} \vec \Phi_n(y)
\int_{\wh \Gamma_k}\Int \mathcal B_2(x;\eta,s) {\rm e}^{\eta \rho -xs + \wh
  V_2(s)-V_2(\eta) } \frac
    {\Amat(z)(z-x)}{(z-\rho)(x-\rho)}\wh \Psi_n(\rho) =\cr 
=&\& \frac 1 {2i\pi} \int_{\Gamma_j}{\rm e}^{zy} \vec \Phi_n(y)
\int_{\wh \Gamma_k}\Int \mathcal B_2(x;\eta,s) {\rm e}^{\eta \rho -xs + \wh
  V_2(s)-V_2(\eta) } \Amat(z)\le(\frac
1{x-\rho} - \frac 1{z-\rho}\ri)\wh \Psi_n(\rho) =\cr 
=&\& \un \Phi_n^{(j)}(z) \Amat(z)\wh \Psi_n^{(k)} (x) -\frac 1{2i\pi}
\sum_{r=0}^{n-1}  \int_{\Gamma_j}{\rm e}^{zy} \phi_r(y) \int_{\wh
  \Gamma_k}\Int \mathcal B_2(x;\eta,s) {\rm e}^{\eta \rho -xs + \wh
  V_2(s)-V_2(\eta) } \psi_r(\rho) =\cr 
\m{=}^{\star}&\& \un \Phi_n^{(j)}(z) \Amat(z) \wh \Psi_n(x) - \frac
1{2i\pi} \int_{\Gamma_j} \int_{\wh\Gamma_k} \mathcal B_2(x;y,s) {\rm
  e}^{yz-xs+\wh V_2(s) -V_2(y)} \ , 
\eea 
where the identity marked $\star$ is valid for $n\geq d_2$ (so that
the kernel reproduces the polynomial $\mathcal B_2(x;\eta,s)$ of
degree $d_2-1$). 

The proof of the second form of ${\bf (b)}$ is only marginally different in that we have
to use the second form of the  principal CDI for the kernel $\wh K_{11}$ (in Thm. \ref{CDI}).
{\bf Q.E.D.}\par \vskip 5pt  
\subsection{Proof of Theorem \ref{ODEdual}}
\label{proofODEdual}
Let $n-1\leq m\leq n+d_2-1$: in the following chain of equalities all
the steps are ``elementary'' and hence the computation is
straightforward. For reader's convenience we have tried to make
annotations on the formula in order to highlight less obvious steps.
\bea
\pa_x \un \phi_m^{(0)}  &\&  =V_1'(x )\un \phi_m^{(0)}+ {\rm e}^{V_1(x)}  \Int {\rm e}^{\xi y}(-\pa_\xi) \frac { {\rm e}^{-V_1(\xi) }\phi_m(y)  } {x-\xi} = \cr
=&\&- \undergroup{\frac {{\rm e}^{V_1(x) - V_1(\xi)}}{x-\xi} \un \phi_m^{(\varkappa)}(\xi)\bigg|_{\xi\in \pa_x \varkappa} }_{=:(B)} +
\undergroup{
{\rm e}^{V_1(x)}  \Int  \frac {(V_1'(x) -V_1'(\xi)){\rm e}^{-V_1(\xi) + \xi y  } \phi_m(y)  } {x-\xi}}_{=:(C)}
    +  {\rm e}^{V_1(x)}  \Int  \frac {y \, {\rm e}^{-V_1(\xi) + \xi y  } \phi_m (y)  } {x-\xi} =\cr
=&\& (-B + C) + \sum_{j=0}^{n+d_2}\un\phi_j^{(0)}(x)  \Int \psi_j(\rho) \eta \phi_m(\eta){\rm e}^{\rho\eta}= \cr 
=&\&  (-B + C) + \sum_{j=0}^{n-1}\un \phi_j^{(0)}(x) \Int \psi_j(\rho) \eta \phi_m(\eta){\rm e}^{\rho\eta}
  + 
  \sum_{j=n}^{n+d_2 } \un\phi_j^{(0)}(x)   \overgroup{\Int \psi_j(\rho) \eta \phi_m(\eta){\rm e}^{\rho\eta}}^{=: P_{jm}}    =\cr 
=&\&
(-B+C)+  \sum_{j=n}^{n+d_2 } \un\phi_j^{(0)}(x)   P_{jm} 
+
 \sum_{j=0}^{n-1}\un \phi_j^{(0)}(x) \le[\psi_j(\rho) \un \phi_m^{(\varkappa)}(\rho) \ri]_{\rho\in \pa_x \varkappa} + 
\cr
&\&-
 \sum_{j=0}^{n-1}\un \phi_j^{(0)}(x) \Int  \phi_m(\eta){\rm e}^{\rho\eta-V_1(\rho)}(\pa_\rho-V_1'(\rho)) \pi_j(\rho) 
=\cr
=&\&
(-B+C)+  \sum_{j=n}^{n+d_2 } \un\phi_j^{(0)}(x)   P_{jm} 
+
  \le[\frac {\un \Phi_n^{(0)}(x)\wh\Amat(x)\wh \Psi_n(\rho) + {\rm
	e}^{V_1(x)-V_1(\rho)}}{x-\rho} \un \phi_m^{(\varkappa)}(\rho) \ri]_{\rho\in \pa_x \varkappa} + 
\cr
&\&+
 \sum_{j=0}^{n-1}\un \phi_j^{(0)}(x) \Int  \phi_m(\eta){\rm e}^{\rho\eta-V_1(\rho)}V_1'(\rho) \pi_j(\rho) 
=\cr
=&\&
(C)+  \sum_{j=n}^{n+d_2 } \un\phi_j^{(0)}(x)   P_{jm} 
+
  \le[\frac {\un \Phi_n^{(0)}(x)\Amat(x) \wh\Psi_n(\rho) } {x-\rho}
    \un \phi_m^{(\varkappa)}(\rho) \ri]_{\rho\in \pa_x \varkappa} + 
\cr
&\&+
\Int  \phi_m(\eta){\rm e}^{\rho\eta}V_1'(\rho)\frac { \un \Phi_n^{(0)}(x)\Amat(x)  \wh  \Psi_n(\rho) + {\rm e}^{V_1(x)-V_1(\rho)}}{x-\rho} 
=\cr
=&\&
  \sum_{j=n}^{n+d_2 } \un\phi_j^{(0)}(x)   P_{jm} 
+
  \le[\frac {\un \Phi_n^{(0)}(x)\Amat(x)\wh\Psi_n(\rho) } {x-\rho} \un
  \phi_m^{(\varkappa)}(\rho) \ri]_{\rho\in \pa_x \varkappa} + 
\cr
&\&+
\Int  \phi_m(\eta){\rm e}^{\rho\eta}V_1'(\rho)\frac { \un \Phi_n^{(0)}(x)\Amat(x)   \wh \Psi_n(\rho)}{x-\rho} + V_1'(x){\rm e}^{V_1(x)}   \Int  \frac {{\rm e}^{-V_1(\xi) + \xi y  } \phi_m(y)  } {x-\xi}
=\cr
=&\&
  \sum_{j=n}^{n+d_2 } \un\phi_j^{(0)}(x)   P_{jm} 
+
  \le[\frac {\un \Phi_n^{(0)}(x)\wh\Psi_n(\rho) } {x-\rho} \un \phi_m^{(\varkappa)}(\rho) \ri]_{\rho\in \pa_x \varkappa} + 
\Int  \phi_m(\eta){\rm e}^{\rho\eta}\frac {V_1'(\rho)-V_1'(x)}{x-\rho}  \un \Phi_n^{(0)}(x)\Amat(x)  \wh  \Psi_n(\rho) 
\cr
&\&+
V_1'(x) \Int  \phi_m(\eta){\rm e}^{\rho\eta} \frac { \un \Phi_n^{(0)}(x)\Amat(x)   \wh \Psi_n(\rho) +{\rm e}^{V_1(x)-V_1(\rho)} }{x-\rho}
=\cr
=&\&
  \sum_{j=n}^{n+d_2 } \un\phi_j^{(0)}(x)   P_{jm} 
+
  \le[\frac {\un \Phi_n^{(0)}(x)\Amat(x) \wh\Psi_n(\rho) } {x-\rho}
  \un \phi_m^{(\varkappa)}(\rho) \ri]_{\rho\in \pa_x \varkappa} -  \un \Phi_n^{(0)}(x)\Amat(x)   \Int  \wh \Psi_n(\rho) 
\phi_m(\eta){\rm e}^{\rho\eta}\frac {V_1'(\rho)-V_1'(x)}{\rho-x} 
\cr
&\&+
V_1'(x)\sum_{j=0}^{n-1} \un\phi_j^{(0)} (x)  \Int  \phi_m(\eta){\rm e}^{\rho\eta} \psi_j(\rho) 
=\cr
=&\&
  \sum_{j=n}^{n+d_2 } \un\phi_j^{(0)}(x)   P_{jm} 
+
  \le[\frac {\un \Phi_n^{(0)}(x)\Amat(x) \wh\Psi_n(\rho) } {x-\rho}
  \un \phi_m^{(\varkappa)}(\rho) \ri]_{\rho\in \pa_x \varkappa} -  \un \Phi_n^{(0)}(x)\Amat(x)   \Int \wh  \Psi_n(\rho) 
\phi_m(\eta){\rm e}^{\rho\eta}\frac {V_1'(\rho)-V_1'(x)}{\rho-x} 
\cr
&\&+
V_1'(x) \un\phi_{n-1}^{(0)} (x) \delta_{m,n-1}
\eea
We note that in this last expression we have $\pa_x \un\phi_m^{(0)}(x)$ expressed purely in terms of $\un\phi_\ell^{(0)}(x)$ for $\ell=n-1,\dots n+d_2$, the value $\ell=n+d_2$ entering only in the first expression. Given that $\un\phi_n^{(0)}(x)$ satisfies the same multiplicative recurrence relations as the Fourier--Laplace transforms for $n\geq 1$, we can re-express $\un\phi_{n+d_2}^{(0)}$ in terms of the elements of the window $\un \Phi_n^{(0)}(x)$, obtaining the result.

The computation for the Fourier-Laplace transforms gives also the same differential equation, indeed
\bea
\pa_x \un\phi_m^{(r)}(x) =&\&\int_{\Gamma_{y,r}} {\rm e}^{xy}\phi_m(y) = 
\sum_{j=0}^{n+d_2} \un\phi_j^{(r)} (x) \Int {\rm e}^{\eta \rho} \eta \phi_m(\eta) \psi_j(\rho) = \cr 
=&\&
\sum_{j=n}^{n+d_2} \un \phi_j^{(r)}(x) P_{jm} + \sum_{j=0}^{n-1} \un
\phi_j^{(r)}(x) \Int  {\rm e}^{\eta \rho-V_1(\rho) } \phi_m(\eta)
(-\pa_\rho + V_1'(\rho) ) \pi_j(\rho) +\cr 
&\&+  \sum_{j=0}^{n-1}\un
\phi_j^{(r)}(x) \le[\psi_j(\rho) \un \phi_m^{(\varkappa)}(\rho) \ri]_{\rho\in \pa_x \varkappa} =\cr
=&\&
\sum_{j=n}^{n+d_2} \un \phi_j^{(r)}(x) P_{jm} + \un
\Phi_n^{(r)}(x)\Amat(x) \Int  {\rm e}^{\eta \rho } \phi_m(\eta)
\frac{V_1'(\rho)}{x-\rho}   \wh \Psi_n(\rho) +\cr 
&\& +  \un
\Phi_n^{(r)}(x)\Amat(x)  \le[\frac {\wh \Psi_n(\rho) \un \phi_m^{(\varkappa)}(\rho)}{x-\rho} \ri]_{\rho\in \pa_x \varkappa} =\cr
=&\&
\sum_{j=n}^{n+d_2} \un \phi_j^{(r)}(x) P_{jm} + \un \Phi_n^{(r)}(x)\Amat(x) \Int   \wh \Psi_n(\rho) {\rm e}^{\eta \rho } \phi_m(\eta) \frac{V_1'(\rho)-V_1'(x) }{x-\rho}   +\cr
&\&+ V_1'(x) \un \Phi_n^{(r)}(x)\Amat(x) \Int  \frac{  \wh\Psi_n(\rho)}{x-\rho}   {\rm e}^{\eta \rho } \phi_m(\eta) 
+ \un \Phi_n^{(r)}(x)\Amat(x)  \le[\frac {\wh \Psi_n(\rho) \un \phi_m^{(\varkappa)}(\rho)}{x-\rho} \ri]_{\rho\in \pa_x \varkappa} =\cr
=&\&
\sum_{j=n}^{n+d_2} \un \phi_j^{(r)}(x) P_{jm} + \un \Phi_n^{(r)}(x)\Amat(x) \Int   \wh \Psi_n(\rho) {\rm e}^{\eta \rho } \phi_m(\eta) \frac{V_1'(\rho)-V_1'(x) }{x-\rho}   +\cr
&\&+ V_1'(x) \sum_{j=0}^{n-1} \un \phi_j^{(r)}(x) \Int    \psi_j(\rho)   {\rm e}^{\eta \rho } \phi_m(\eta) 
+ \un \Phi_n^{(r)}(x)\Amat(x)  \le[\frac {\wh \Psi_n(\rho) \un \phi_m^{(\varkappa)}(\rho)}{x-\rho} \ri]_{\rho\in \pa_x \varkappa} =\cr
=&\&
\sum_{j=n}^{n+d_2} \un \phi_j^{(r)}(x) P_{jm} + \un \Phi_n^{(r)}(x)\Amat(x) \Int   \wh \Psi_n(\rho) {\rm e}^{\eta \rho } \phi_m(\eta) \frac{V_1'(\rho)-V_1'(x) }{x-\rho}   +\cr
&\&+ V_1'(x) \delta_{m,n-1}  \un \phi_m^{(r)}(x) + \un
\Phi_n^{(r)}(x)\Amat(x)  \le[\frac {\wh \Psi_n(\rho)\un \phi_m^{(\varkappa)}(\rho)}{x-\rho} \ri]_{\rho\in \pa_x \varkappa} \ .
\eea
The coefficients of these expressions in terms of $\un \phi_{n-1},\dots \un \phi_{n+d_2-1}$ are precisely the same as for the previous computation, hence completing the proof. 
{\bf Q.E.D.}\par \vskip 5pt
\subsection{Proof of Theorem \ref{ODEdirhat}}
\label{proofODEdirhat}
Let $n-d_2\leq m \leq n$ and let us compute
\bea
\pa_x \wh \psi_m(x) =&\& {\rm e}^{-V_1(x) } (\pa_x - V_1'(x)) \wh \pi_m(x) =-V_1'(x)\wh \psi_m(x) +   \sum_{j=0}^{n-1}\wh \psi_j (x) \Int\wh  \pi_m'(\xi ) {\rm e}^{\xi y - V_1(\xi)}\wh  \phi_j(y) =\cr 
=&\&
-V_1'(x)\wh \psi_m(x) +   \sum_{j=0}^{n-1}\wh \psi_j (x) \le[ \wh \psi_m(\xi) \un{\wh \phi}_j^{(\varkappa)}(\xi) \ri]_{\xi\in \pa_x \varkappa}   - \sum_{j=0}^{n-1}\wh \psi_j(x) \Int \wh \psi_m(\xi ) {\rm e}^{\xi y} y  \wh \phi_j(y) 
=\cr
=&\&
-V_1'(x) \wh \psi_m(x) + \le[ \frac{\wh \psi_m(\xi) \un \Phi _n^{(\varkappa)}(\xi)}{\xi-x}  \ri]_{\xi\in \pa_x \varkappa}  \Amat(x)\wh \Psi_j (x)    - \sum_{j=0}^{n-1}\wh \psi_j(x)  \Int \wh \psi_m(\xi ) {\rm e}^{\xi y} (y -V_1'(\rho))\wh  \phi_j(y) 
=\cr
=&\&
-V_1'(x) \wh \psi_m(x) + \le[ \frac{\wh \psi_m(\xi) \un \Phi _n^{(\varkappa)}(\xi)}{\xi-x}  \ri]_{\xi\in \pa_x \varkappa}  \Amat(x)\wh \Psi_j (x)  +
 \sum_{j=0}^{n-1}\wh \psi_j(x)  \Int \wh \psi_m(\xi ) {\rm e}^{\xi y} V_1'(\rho)\wh  \phi_j(y) +\cr 
 &\&
  - \sum_{j={\bf m-1}}^{n-1}\wh \psi_j(x)  \Int \wh \psi_m(\xi ) {\rm e}^{\xi y} y \wh  \phi_j(y) 
=\cr
=&\&
-V_1'(x) \delta_{mn} \wh \psi_n(x) + \le[ \frac{\wh \psi_m(\xi) \un \Phi _n^{(\varkappa)}(\xi)}{\xi-x}  \ri]_{\xi\in \pa_x \varkappa}  \Amat(x)\wh \Psi_j (x)  +\cr
&\& + 
 \sum_{j=0}^{n-1}\wh \psi_j(x)  \Int \wh \psi_m(\xi ) {\rm e}^{\xi y}( V_1'(\rho)-V_1'(x))\wh  \phi_j(y) 
  - \sum_{j= m-1}^{n-1}\wh \psi_j(x)  \wh P_{mj}
=\cr
=&\&
-V_1'(x) \delta_{mn} \wh \psi_n(x) + \le[ \frac{\wh \psi_m(\xi) \un \Phi _n^{(\varkappa)}(\xi)}{\xi-x}  \ri]_{\xi\in \pa_x \varkappa}  \Amat(x)\wh \Psi_j (x)  +\cr 
&\& + 
  \Int \wh \psi_m(\xi ) {\rm e}^{\xi y}\frac{ V_1'(\rho)-V_1'(x)}{\rho-x}   \vec \Phi_n(y) \Amat (x) \wh \Psi_j(x)  
  - \sum_{j= m-1}^{n-1}\wh \psi_j(x) \wh P_{mj}\ .
\eea
The last term contains $\wh \psi_{n-d_2-1}$ (for $m=n-d_2$) which is
"outside" of the window of the quasipolynomials. Using the recurrence
relations and re-expressing it in terms of elements in the window
(using the ladder matrices) we obtain the formula.  

For completeness one should also consider the other columns of the
fundamental system $\boldsymbol {\wh \Psi}_n$  and show that they
satisfy the same differential relation as the quasipolynomials. Let
$n-d_2\leq m \leq n$, then  
\bea
\pa_x \wh \psi_m^{(r)} =&\& \frac 1{2i\pi}  \pa_x \int_{\wh
  \Gamma_{y,r}} \Int \mathcal B_2(x;y,s) {\rm e}^{\xi y-xs+ \wh V_2(s)
  - V_2(y)} \frac {\wh \psi_m(\xi)}{x-\xi}=\cr 
=&\&  \frac 1{2i\pi}\int_{\wh \Gamma_{y,r}} \Int {\rm e}^{\xi y-xs+
  \wh V_2(s) - V_2(y)}\wh \psi_m(\xi) (\pa_x -s) \frac {\mathcal
  B_2(x;y,s)}{x-\xi}= \cr 
=&\&  \frac 1{2i\pi}\int_{\wh \Gamma_{y,r}} \Int {\rm e}^{\xi y-xs+ \wh V_2(s) - V_2(y)}\wh \psi_m(\xi)\le[
\frac {B_2(y)-B_2(s)}{(y-s)(x-\xi)} - s\frac {\mathcal B_2(x;y,s)}{(x-\xi)} - \mathcal B_2(x;y,s) \pa_\xi\frac 1{x-\xi}
\ri]=\cr 
=&\&- \overgroup{
\frac 1{2i\pi} \int_{\wh \Gamma_{y,r}} \Int \pa_\xi\le( \mathcal
B_2(x;y,s)    {\rm e}^{\xi y-xs+ \wh V_2(s) - V_2(y)}\frac {\wh
  \psi_m(\xi)}{x-\xi} \ri)        
}^{=:(B)} +\cr
&\& + \int_{\wh \Gamma_{y,r}} \Int \frac {\wh \psi_m(\xi)}{x-\xi} {\rm
  e}^{\xi y-xs+ \wh V_2(s) - V_2(y)} \le[ \frac {B_2(y)-B_2(s)}{y-s} +
  (y-s) \mathcal B_2(x;y,s) \ri]  + \cr 
&\& + \int_{\wh \Gamma_{y,r}} \Int \mathcal B_2(x;y,s){\rm e}^{\xi
  y-xs+ \wh V_2(s) - V_2(y)}  \frac {(\pa_\xi-V_1'(\xi))\wh
  p_m(\xi)}{x-\xi} = \cr 
&\& = -(B) + \int_{\wh \Gamma_{y,r}} {\rm e}^{\wh V_2(s)-xs}
\overgroup{\Int \wh \psi_m(\xi) {\rm e}^{-V_2(y) + \xi y} B_2(y)
}^{ = \sqrt{\wh h_0} \delta_{m0}\scriptsize{ \hbox{ because } } \wh V_2 =
  V_2-\ln B_2 } +\cr
&\& - 
\overgroup{
\frac 1{2i\pi} \int_{\wh \Gamma_{y,r}} \Int  \mathcal B_2(x;y,s)
      {\rm e}^{\xi y-xs+ \wh V_2(s) - V_2(y)}\frac {V_1'(\xi) \wh
	\psi_m(\xi)}{x-\xi}       
}^{=:(C)}  + \cr
&\& + \int_{\wh \Gamma_{y,r}} \Int \mathcal B_2(x;y,s){\rm e}^{\xi
  y-xs+ \wh V_2(s) - V_2(y)}  \frac {\wh p_m'(\xi)}{x-\xi}=\cr 
=&\& -(B)-(C)+\overgroup{ \sqrt{\wh h_0} \delta_{m0} \int_{\wh \Gamma_{y,r}} {\rm e}^{\wh V_2(s)-xs}}^{=:(D)} +  \sum_{j=0}^{n-1} \wh \psi_{j}^{(r)}(x) \Int \wh p_m\,' (\xi)\wh \phi_j(y){\rm e}^{-V_1(\xi)+\xi y} =\cr
=&\& -(B)-(C) + (D)+\sum_{j=0}^{n-1} \wh\psi_j^{(r)}(x)\le[ \wh
  \psi_m(\xi) \un {\wh \phi}^{(\varkappa)}_j(\xi)\ri]_{\xi\in \pa_x
  \varkappa} + \cr 
  &\& + \sum_{j=0}^{n-1} \wh \psi_j^{(r)}(x) \Int (V_1'(\xi) -
V_1'(x)) \wh \psi_m(\xi)\wh\phi_j(y) {\rm e}^{\xi y} +\cr 
&\& +  V_1'(x) (1-\delta_{m,n}) \psi_m^{(r)}(x)  - \sum_{j=0}^{n-1}
  \wh \psi_j^{(r)}(x) \Int \wh \psi_m(\xi) y\,\wh \phi_j(y){\rm
  e}^{\xi y} = \cr 
\m{=}^{\scriptsize \hbox{[aux CDI]}} &\& -(B)-(C) + \frac {\wh
  \psi_m(\xi) \un {\wh \Phi}_n^{(\varkappa)} (\xi)}{\xi - x}
\bigg|_{\xi\in \pa_x \varkappa} \Amat(x) \wh \Psi_n(x)   + \cr
&\& \overgroup{- \frac
  1{2i\pi} \int_{\wh \Gamma_{y,r}} \Int \pa_\xi \le( \frac {\wh
  \psi_m(\xi)}{\xi-x}  \mathcal B_2(x;y,s) {\rm e}^{y\xi - xs +
  \wh V_2(s) - V_2(y)} \ri) }^{= (B)}+ \cr
&\&
 +
\Int \frac {V_1'(\xi)-V_1'(x)}{\xi-x}  {\rm e}^{\xi y}  \wh \psi_m(\xi)
\vec \Phi_n(y) \Amat(x) \wh \Psi_n^{(r)}(x)  + \cr 
&\&
-\frac 1{2i\pi} \int_{\wh \Gamma_{y,r}} \Int
  (\m{V_1'(\xi)}^ {\ds \m{\uparrow} ^{[{\rm
  gives\ } (C)]}} -V_1'(x)) \frac
  {\wh  \psi_m(\xi)}{\xi-x} {\rm e}^{\xi y-xs + \wh V_2(s) - V_2(y)} 
+\cr
&\& +  V_1'(x) (1-\delta_{m,n)} \psi_m^{(r)}(x)  -
\sum_{j=m-1}^{n-1}\wh P_{mj}
\wh \psi_j^{(r)}(x)   =  \cr
=&\&  \frac {\wh  \psi_m(\xi) \un {\wh \Phi}_n^{(\varkappa)} (\xi)}{\xi - x}
\bigg|_{\xi\in \pa_x \varkappa} \Amat(x) \wh \Psi_n(x) + 
\Int \frac {V_1'(\xi)-V_1'(x)}{\xi-x}  {\rm e}^{\xi y}  \wh \psi_m(\xi)
\vec \Phi_n(y) \Amat(x) \wh \Psi_n^{(r)}(x)  \cr
&\& -\delta_{mn}   V_1'(x)  \psi_m^{(r)}(x)  - \sum_{j=0}^{n-1}
\wh \psi_j^{(r)}(x) \Int \wh \psi_m(\xi) y\,\wh \phi_j(y){\rm e}^{\xi
  y} 
 \eea
This is the same expression as for the quasipolynomials: since the
auxiliary wave functions $\wh \psi_j^{(r)}(x)$ satisfy the same
multiplicative recurrence relation (for $n$ large enough) as the
quasipolynomials, re-expressing $\wh \psi_{n+1}^{(r)}(x)$ in terms of
the elements of the window yields the same differential equation.
{\bf Q.E.D.}\par \vskip 5pt

\subsection{Proof of Theorem \ref{perfect}}
\label{proofperfect}
For brevity we denote $\Amat_n(x)$ simply by $\Amat(x)$ during this proof.
Since the rows (columns) of $\un{\mathbf \Phi_n}$ ($\wh{\mathbf \Psi}_n$) are of two types, we need to carry out four types of computations
\bea
& {\bf(a)} = \un \Phi_n^{(0)} (x)\Amat(x) \wh {\Psi}_n^{(0)} (x)\ , & 
{\bf (b)}  = \un \Phi_n^{(0)} (x)\Amat(x) \wh {\Psi}_n^{(j)} (x)\ ,\ j=1\dots d_2\cr
&{\bf(c)} = \un \Phi_n^{(j)} (x)\Amat (x)\wh {\Psi}_n^{(0)} (x)\ ,\ j=1\dots d_2 &
{\bf (d)}  = \un \Phi_n^{(\ell)} (x)\Amat (x) \wh {\Psi}_n^{(m)} (x)\ ,\ \ell,m =1\dots d_2
\eea
It follows trivially from (\ref{regCDI}) that ${\bf (c)} = 0$ (set $x=x'$ in the LHS).
For ${\bf (a)}$ we have 
\bea
{\bf (a)}&=&   {\rm e}^{V_1(x)} 
\int\!\!\! \int_\varkappa \frac{\Phi_n(\xi)}{x-\zeta} {\rm
  e}^{-V_1(\zeta)+\xi\zeta} \Amat(x) \wh \Psi_n(x) =\\
&&=
{\rm e}^{V_1(x)}\int\!\!\!   \int_\varkappa\!\!\!{\rm d}\zeta\, {\rm e}^{-V_1(\zeta)+\xi\zeta}\sum_{j=0}^{n-1}\wh \phi_j(\xi)
 \wh \psi_j(x) =
{\rm e}^{V_1(x)}\int\!\!\!   \int_\varkappa\!\!\!{\rm d}\zeta\, {\rm e}^{-V_1(\zeta) +\xi\zeta} \wh\phi_0(\xi)
 \wh \psi_0(x) =1 \nonumber
\eea
where we have used that $ \wh \phi_j(\zeta)$, $j\geq 1$ are  orthogonal to
$p(\xi)\equiv 1$. Note also that we had to use the CDI in the form (\ref{hatCDI}).
Then we have to compute for $1\leq \ell,m\leq d_2$  (we suppress
explicit reference to the variables of integration because there is no
possibility of ambiguity)
\bea
{\bf (d)} & =& \frac 1{2i\pi}
\un \Phi_n^{(\ell)}(x) \int\!\!\!\int_\varkappa\int_{s\in\wh \Gamma_m}\mathcal B_2(x; \eta,s) \frac{\Amat(x)\wh \Psi_n(\xi)}{x-\xi} {\rm
	    e}^{\xi\eta-xs - V_2(\eta)+ \wh  V_2(s)}=\\
&=&\frac 1{2i\pi}\sum_{j=0}^{n-1} \un{ \phi}_j^{(\ell)}(x)  \int\!\!\!\int_\varkappa\int_{s\in\wh \Gamma_m} \mathcal B_2(x;\eta,s) { \psi_j(\xi)}{\rm
	    e}^{\xi\eta-xs - V_2(\eta)+  \wh V_2(s)}=\\
&=&\frac 1{2i\pi}\sum_{j=0}^{n-1} \int_{\wh \Gamma_m} \!\!\!{\rm d}s
	  \int_{\Gamma_\ell}\!\!\!{\rm d}y\, \phi_j(y){\rm e}^{xy} 
\int\!\!\!\int_\varkappa\mathcal B_2(x;\eta,s) {\psi_j(\xi)}{\rm
	    e}^{\xi\eta-xs - V_2(\eta)+ \wh  V_2(s)}\stackrel{\star}{=}\\
&=&\frac 1{2i\pi} \int_{\wh \Gamma_m} \!\!\!{\rm d}s \int_{\Gamma_\ell}\!\!\!{\rm d}y\, 
\mathcal B_2(x;y,s)
	  {\rm e}^{x(y-s)- V_2(y)+ \wh V_2(s)} = \\
	  &=& \mathcal B_2(\Gamma_{y,\ell},\wh\Gamma_{y,m}) = \delta_{\ell m}\ ,
\eea
where in the step marked with a star we have used that for the
polynomial of $\eta$ $P(\eta):=\mathcal B_2(x;\eta,s)$ is
reproduced by the kernel
\be
P(y) = \sum_{j=0}^{n-1} s_j(y)\int\!\!\!\int_{\varkappa}{\rm
  d}\eta {\rm d}\xi  \psi_j(\xi){\rm e}^{-V_2(\eta)+\xi\eta} P(\eta)
\ee
provided that $n-1\geq {\rm deg}P = d_2-1$. Note also that in this latter computation we are forced to use the other form of the CDI (\ref{regCDI}).
%
%
Finally we need to compute ${\bf (b)}$, which involves quintuple integrals 
\bea
{\bf (b)} &\&=\frac{ {\rm e}^{V_1(x)} }{2i\pi}
 \int\!\!\!\int_\varkappa \frac{\Phi_n(\rho)}{x-\zeta} {\rm
  e}^{-V_1(\zeta)+\rho\zeta} 
 \int\!\!\!\int_\varkappa\int_{s\in \wh \Gamma_m} 
 \mathcal B_2(x;\eta,s)\frac{\Amat(x) \wh \Psi_n(\xi)}{x-\xi} {\rm
	    e}^{\xi\eta-xs - V_2(\eta)+ \wh  V_2(s)}=\nonumber\\    
&&=\frac{ {\rm e}^{V_1(x)} }{2i\pi}
 \int\!\!\!\int_\varkappa \frac{\Phi_n(\rho )}{x-\zeta} {\rm
  e}^{-V_1(\zeta)+\rho\zeta} 
 \int\!\!\!\int_\varkappa\int_{s\in \wh \Gamma_m} 
 \mathcal B_2(x;\eta,s)\frac{\Amat(\zeta) \wh \Psi_n(\xi)}{x-\xi} {\rm
	    e}^{\xi \eta-xs - V_2(\eta)+ \wh  V_2(s)} + \\
&& +\frac{ {\rm e}^{V_1(x)} }{2i\pi}
\underbrace{\Int {\Phi_n(\rho)}[\wh L,p_n]
  {\rm
  e}^{-V_1(\zeta)+\rho\zeta} }_{=0 \hbox { if } n\geq q_1}
 \int\!\!\!\int_\varkappa\int_{s\in \wh \Gamma_m} 
 \mathcal B_2(x;\eta,s)\frac{ \wh \Psi_n(\xi)}{x-\xi} {\rm
	    e}^{\xi\eta-xs - V_2(\eta)+ \wh  V_2(s)} =\nonumber\\
&&=\frac{ {\rm e}^{V_1(x)} }{2i\pi} \sum_{j=0}^{n-1}
	  \int\!\!\!\int_\varkappa
	  \!\!\!\int\!\!\!\int_\varkappa\int_{s \in \check
  \Gamma_m}\!\!\!{\rm d}s\, {\rm e}^{-V_1(\zeta) + \zeta \rho + \xi\eta -x s
	    - V_2(\eta)+ \wh V_2(s)}  \phi_j(\rho) \psi_j(\xi) \mathcal B_2(x;\eta,s)\frac
	  {\zeta-\xi}{(x-\zeta)(x-\xi)} =\nonumber \\ 
&&=\frac{ {\rm e}^{V_1(x)} }{2i\pi} \sum_{j=0}^{n-1} \int\!\!\!{\rm
	    d}y\int\!\!\!\int_\varkappa\!\!\!\int\!\!\!\int_\varkappa\int_{s\in \wh \Gamma_m}\!\!\!{\rm d}s\, {\rm e}^{-V_1(\zeta)
 +  \xi\eta -x s 
	    - V_2(\eta)+  \wh V_2(s)+\zeta \rho}  \phi_j(\rho ) \psi_j(\xi)  \mathcal B_2(x;\eta,s)
	    \left( \frac{1}{x-\zeta}-\frac
	  1{x-\xi}\ri) =\nonumber  \\
&&= \frac{ {\rm e}^{V_1(x)} }{2i\pi} \sum_{j=0}^{n-1} \int\!\!\!{\rm d}y\int\!\!\!\int\!\!\!\int\!\!\!\int_{\check
  \Gamma}\!\!\!{\rm d}s \,{\rm e}^{-V_1(\zeta) + \xi\eta -x s
	    - V_2(\eta)+  \wh V_2(s)+\zeta \rho}  \phi_j(\rho) \psi_j(\xi)  \mathcal B_2(x;\eta,s) \frac{1}{x-\zeta}  +\nonumber  \\
&&\hspace{1cm}- \frac{ {\rm e}^{V_1(x)} }{2i\pi} \int\!\!\!\int\!\!\!\int_{\check
  \Gamma}\!\!\!{\rm d}s \,  \mathcal B_2(x;\eta,s) \frac{1}{x-\xi} {\rm e}^{ \xi\eta -x s
	    - V_2(\eta)+ \wh V_2(s)-V_1(\xi)} =\nonumber \\ 
&&= \frac{ {\rm e}^{V_1(x)} }{2i\pi}\int\!\!\!{\rm d}y\int\!\!\!\int_{\check
  \Gamma}\!\!\!{\rm d}s\, {\rm e}^{-V_1(\zeta) - V_2(\rho) +\zeta \rho -x s
	    + \wh  V_2(s) }   \mathcal B_2(x;\rho,s) \frac{1}{x-\zeta}  +\nonumber  \\
&&\hspace{1cm}-\frac{ {\rm e}^{V_1(x)} }{2i\pi}\int\!\!\!\int\!\!\!\int_{\check
  \Gamma}\!\!\!{\rm d}s \,{\rm e}^{ -V_1(\xi)- V_2(\eta)+ \xi\eta -x s
	    +  \wh V_2(s)} \mathcal B_2(x;\eta,s) \frac{1}{x-\xi}  \equiv 0
\eea
Once more, we are forced to use the CDI in the form (\ref{regCDI}).
This concludes the proof. {\bf Q.E.D.}\par\vskip 5pt


\begin{thebibliography}{99}

\bibitem{AvM} M. Adler and P. Van Moerbeke, ``The Spectrum of Coupled Random
Matrices'', {\it Ann. Math.} {\bf 149}, 921--976 (1999).
%
\bibitem {AvM2} M. Adler and P. Van Moerbeke, 
 ``String-orthogonal polynomials,
string equations and 2-Toda symmetries'', {\it Comm. Pure and Appl. Math. J.},
{\bf 50} 241-290 (1997).
%
\bibitem{jat} M. Bertola, ``Bilinear semi--classical moment
functionals and their integral representation'',
  {\em J. App. Theory}  {\bf 121},  71-99 (2003).
%
\bibitem{BEH} M. Bertola, B. Eynard and J. Harnad, ``Duality,
Biorthogonal Polynomials and Multi--Matrix Models'',
Commun. Math. Phys. {\bf 229}, 73--120 (2002).
%
 \bibitem {BEH2}  M. Bertola, B. Eynard and J. Harnad,
``Differential systems for biorthogonal polynomials appearing in
2-matrix models and the associated Riemann-Hilbert problem'',
Commun. Math. Phys. {\bf 243}, 193--240 (2003).
%
\bibitem{BEH3}  M. Bertola, B. Eynard and  J. Harnad, ``Duality of 
spectral curves arising in two-matrix models'' 
{\it  Theor. Math. Phys.} {\it  Theor. Math. Phys.}  {\bf 134}, 27-38  (2003). 
%
\bibitem{BE} M. Bertola and B. Eynard, ``The PDEs of biorthogonal polynomials 
arising in the two-matrix model'', Saclay preprint SPT-03/139 (2003), nlin.SI/0311033.
%
\bibitem {BHI} M. Bertola, J. Harnad, A. Its, "Dual Riemann--Hilbert  approeach to biorthogonal polynomials".
%
\bibitem{BEHiso}M. Bertola, B. Eynard, J. Harnad, "Semiclassical orthogonal polynomials, matrix models and isomonodromic tau functions", nlin.SI/0410043, Comm. Math. Phys. to appear.
%
\bibitem{BeMo} M. Bertola, M. Y. Mo, "Isomonodromic deformation of resonant rational connections",
nlin.SI/0510011.
%
\bibitem{BeGe}M. Bertola, M. Gekhtman,  "Biorthogonal Laurent polynomials, Toeplitz determinants, minimal Toda orbits and isomonodromic tau functions", nlin.SI/0503050.
%
%
\bibitem{ercoken} N. Ercolani, K. T. R. McLaughlin, ``Asymptotics and integrable structures for biorthogonal polynomials associated to a random two-matrix model'', Physica D: Nonlinear Phenomena, Vol. 152-153, 232--268 (2001).

\bibitem{marcellan} F. Marcell\'an, I. A. Rocha, ``Complex Path Integral
Representation for Semiclassical Linear Functionals'', J. Appr. Theory
{\bf 94}, 107--127, (1998).
%
%
\bibitem{eynardmehta} B. Eynard, M.L. Mehta, ``Matrices coupled in a chain:
eigenvalue correlations'', {\em J. Phys. A: Math. Gen.} {\bf 31}, 4449 (1998),
%
\bibitem{FIK} A. Fokas, A. Its, A. Kitaev, ``The isomonodromy approach
to matrix models in 2D quantum gravity'', {\em Commun. Math. Phys.}
{\bf 147}, 395--430 (1992).
%
\bibitem{ince} E. L. Ince,  "Ordinary Differential Equations", Dover Publications, N.Y. (1944).
%
\bibitem{JMU}  M. Jimbo, T. Miwa and K. Ueno, 
``Monodromy Preserving 
Deformation of Linear Ordinary Differential Equations with Rational
Coefficients I.'', 
{\it Physica} {\bf 2D}, 306-352 (1981).
%
\bibitem{Kap} A. A. Kapaev, ``The Riemann--Hilbert problem for the
bi-orthogonal polynomials'', J.Phys. A {\bf 36} 4629--4640 (2003).
%
%
\bibitem{KM}  A. B. J. Kuijlaars and K. T-R McLaughlin
``A Riemann-Hilbert problem for biorthogonal polynomials'',  math.CV/0310204 
%
\bibitem{shapiromiller} K. S. Miller, H. S. Shapiro, ``On the Linear
Independence of Laplace Integral Solutions of Certain Differential
Equations'', Comm. Pure Appl. Math. {\bf 14} 125--135 (1961).
%
\bibitem{razvan} R. Teodorescu, E. Bettelheim, O. Agam, A. Zabrodin,
  P. Wiegmann, ``Normal random matrix ensemble as a growth problem'', 
Nucl. Phys. B {\bf 700} 521, (2004).
%
 \bibitem{UT} K. Ueno and K. Takasaki, ``Toda Lattice Hierarchy'',
{\it Adv. Studies Pure Math.} {\bf 4}, 1--95 (1984).
%
\end{thebibliography}
\end{document}